\documentstyle[12pt,epsfig]{article}

\newcommand{\goto}{\rightarrow}
\newcommand{\calB}{\mbox{${\cal B}$}}

\newcommand{\etapr}{{\eta^{\prime}}}

\newcommand{\BetapKp}{\mbox{$\calB(B^+\goto\eta^\prime K^+)$}}

\newcommand{\omegak}{\mbox{$\omega K^+$}}
\newcommand{\omegakz}{\mbox{$\omega K^0$}}
\newcommand{\omegapi}{\mbox{$\omega\pi^+$}}
\newcommand{\omegapiz}{\mbox{$\omega\pi^0$}}
\newcommand{\omegah}{\mbox{$\omega h^+$}}

\newcommand{\omegakstz}{\mbox{$\omega K^{*0}$}}

\newcommand{\omegakstp}{\mbox{$\omega K^{*+}$}}

\newcommand{\omegarhoz}{\mbox{$\omega \rho^0$}}
\newcommand{\omegarhop}{\mbox{$\omega \rho^+$}}

\newcommand{\phik}{\mbox{$\phi K^+$}}
\newcommand{\phikz}{\mbox{$\phi K^0$}}
\newcommand{\phipi}{\mbox{$\phi\pi^+$}}
\newcommand{\phipiz}{\mbox{$\phi\pi^0$}}

\newcommand{\rhozpi}{\mbox{$\rho^0 \pi^+$}}
\newcommand{\rhozk}{\mbox{$\rho^0 K^+$}}
\newcommand{\rhozpiz}{\mbox{$\rho^0 \pi^0$}}
\newcommand{\rhozkz}{\mbox{$\rho^0 K^0$}}
\newcommand{\rhompi}{\mbox{$\rho^- \pi^+$}}
\newcommand{\rhomk}{\mbox{$\rho^- K^+$}}

\newcommand{\kstzpiz}{\mbox{$K^{*0} \pi^0$}}
\newcommand{\kstzpi}{\mbox{$K^{*0} \pi^+$}}
\newcommand{\kstzkkp}{\mbox{$K^{*0}_{K^+\pi^-} K^+$}}
\newcommand{\kstppikz}{\mbox{$K^{*+}_{K^0\pi^+} \pi^-$}}
\newcommand{\kstppikp}{\mbox{$K^{*+}_{K^+\pi^0} \pi^-$}}
\newcommand{\kstppi}{\mbox{$K^{*+} \pi^-$}}
\newcommand{\kstpkkz}{\mbox{$K^{*+}_{K^0\pi^+} K^-$}}
\newcommand{\kstpkkp}{\mbox{$K^{*+}_{K^+\pi^0} K^-$}}
\newcommand{\kstpk}{\mbox{$K^{*+} K^-$}}

\def\Title#1{\begin{center} {\Large {\bf #1} } \end{center}}

\begin{document}

\Title{Heavy Quark Decays}

\bigskip\bigskip


\begin{raggedright}  

{\it Ronald A. Poling\index{Poling, R.}\\
School of Physics and Astronomy\\
University of Minnesota\\
Minneapolis, MN 55455}
\bigskip\bigskip
\end{raggedright}

\section{Introduction}

Heavy quark decays are central to the international
effort to test the Standard Model, and the $b$ quark has 
emerged as the focus of this program.  These studies
include detailed investigations of semileptonic and hadronic decays,
as well as increasingly sensitive measurements of rare decays. With
major new $b$-physics initiatives getting under way at nearly all
high energy physics labs, the prospects for definitive tests of the
Standard Model, or the discovery of physics beyond it, are excellent.

\begin{figure}[hb]
\begin{center}
\leavevmode
{\epsfxsize=2.25truein \epsfysize=1.1truein \epsfbox{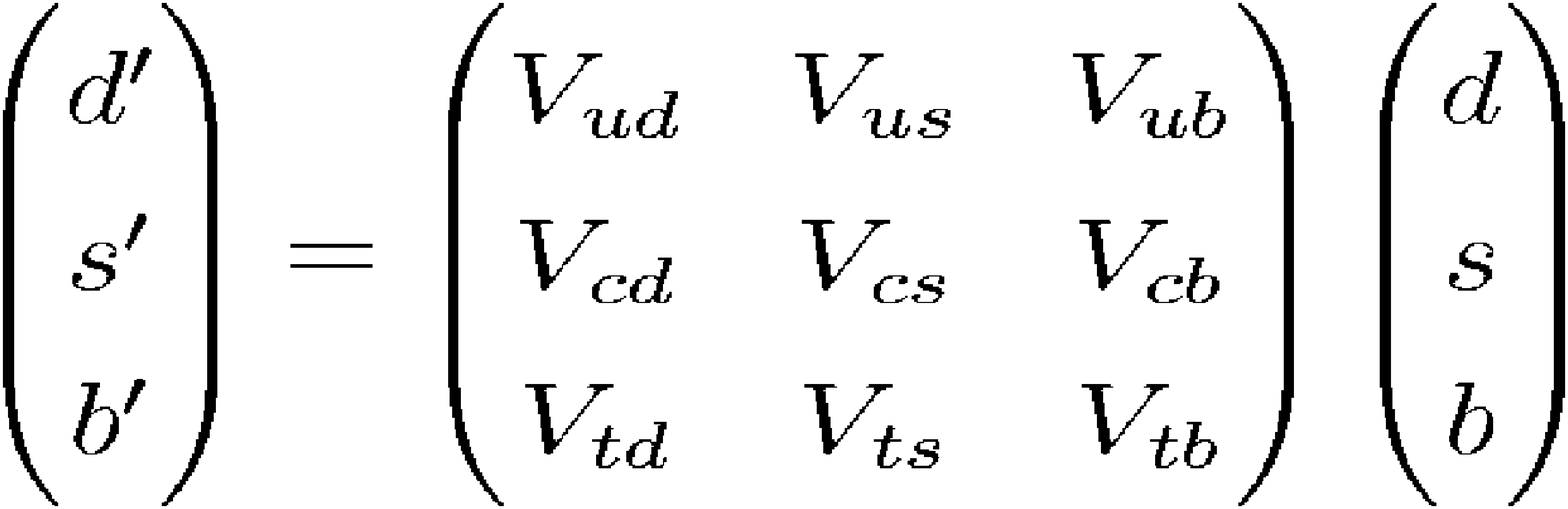}}\hspace*{0.5in}
{\epsfxsize=3.truein \epsfysize=1.truein \epsfbox{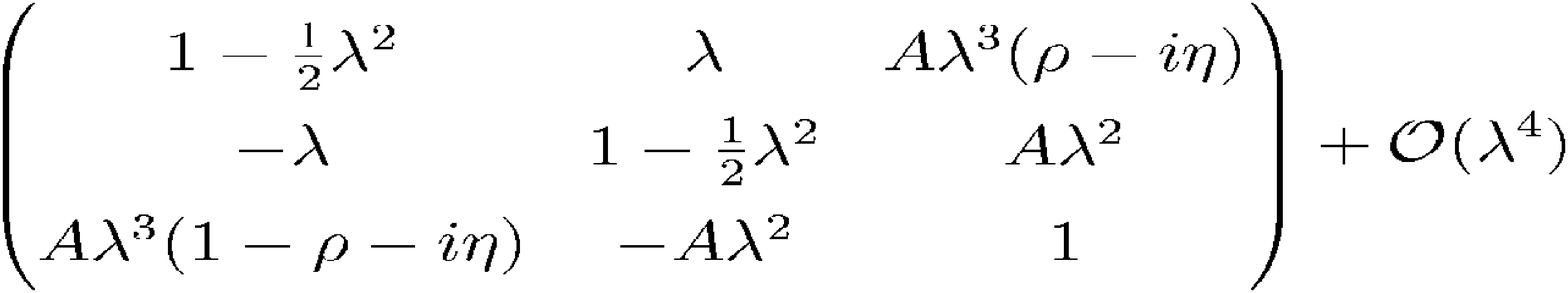}}
\end{center}
\caption{CKM matrix and Wolfenstein parameterization.}
\label{fig:CKM_matrix}
\end{figure}

Flavor physics is interesting because the weak eigenstates of the
quarks are mixtures of the mass eigenstates.  With three generations
the mixing is described by the Cabibbo-Kobayashi-Maskawa
matrix \cite{CKM} (Fig.~\ref{fig:CKM_matrix}).  
Unitarity and the arbitrariness of phases allows the nine complex 
elements to be reduced to four parameters, as most familiarly 
parameterized by Wolfenstein \cite{Wolfenstein:1983yz}.  These
parameters cannot be predicted, and their determination is 
one of our most practical needs.  Furthermore, redundant measurements 
provide powerful tests of the validity of the Standard Model.

Specific measurements include direct determinations of the magnitudes of the
CKM parameters in a variety of processes, and detailed studies of CP 
violation, principally in $s$ and $b$ decays.  ``Overconstraining'' 
the matrix thus is a matter of measuring the lengths of the sides
of the unitarity triangle, as well as its angles,
$\alpha = arg[-V_{td}V^*_{tb} / V_{ud}V^*_{ub}]$, 
$\beta  = arg[-V_{cd}V^*_{cb} / V_{td}V^*_{tb}]$ and
$\gamma = arg[-V_{ud}V^*_{ub} / V_{cd}V^*_{cb}]$. 
We already know quite a bit: $\lambda \simeq 0.22$, $A \simeq 0.8$,
and bounds on $\rho$ and $\eta$ from past measurements.  We urgently 
need precise determinations.

Another powerful probe 
of the limits of the Standard Model 
is provided by rare decays, especially rare $b$ decays.  There are many
observables and many challenging measurements.  They require
very large data samples and mastery of strong-interaction effects that 
obscure our view of the underlying electroweak physics.

The objective of this review is to report some of 
the recent developments in heavy-quark decays, hopefully painting
a picture of our overall state of knowledge and the  
pressing open questions.  Not included are 
the crucial topics of lifetimes and mixing, covered 
elsewhere in these proceedings \cite{Blaylock:1999pz}.  In 
Section~\ref{sec:BSL} I describe the
current status of semileptonic $B$ decays and the determination of
the CKM parameters $|V_{cb}|$ and $|V_{ub}|$.  The focus of 
Section~\ref{sec:RareB} is rare charmless decays, both two-body
hadronic decays and $b \rightarrow s \gamma$.  Section~\ref{sec:CKM}
addresses the interpretation of the various results and implications
for the CKM matrix.  In Section~\ref{sec:Other} I mention a few results
and near-term prospects in charm physics.
This review ends in Section~\ref{sec:Concl} with a brief summary and 
a survey of the outlook for the not-too-distant future in.

The roster of experimental players in this business is growing
with the first operation of several new facilities:
KEK-B/BELLE, PEP-II/BaBar, CESR/CLEO~III, and HERA-B.  
Many recent advances in $b$ physics have been made by the CLEO
experiment working with $B$ mesons just above threshold at the
$\Upsilon(4S)$ resonance.  CLEO has two distinct data samples: 
3.3 million $B {\bar B}$ events in the original CLEO~II detector 
and 6.4 million events obtained since 1996 with CLEO~II.V, upgraded
to include a silicon vertex detector and other improvements.  The 
data sample for CLEO~II.V exceeded the project goal as a result of 
the excellent performance of the CESR storage ring, which reached a 
luminosity of 
$0.8 \times 10^{33}$~cm$^{-2}$s$^{-1}$ by the end of the run.  
The ALEPH, DELPHI, L3, and OPAL experiments at LEP and the SLD 
experiment at the SLC, have investigated faster-moving $B$'s produced 
in $Z^0$ decays.  Each LEP experiment collected roughly
0.9 million $b {\bar b}$ pairs.  With dramatically improved 
SLC performance toward the end of its run, SLD was able to obtain
about 100 thousand $b {\bar b}$'s, with the extra advantages
of polarized beams and outstanding vertexing.  During 
Run~I the Tevatron experiments D0 and, especially, CDF demonstrated
that forefront $b$ physics can be done in a $p {\bar p}$ environment.
CDF's 100~pb$^{-1}$ sample, clean lepton triggers and ability to
tag displaced $B$ vertices produced competitive
measurements not just of lifetimes and mixing, but also of some
rare $B$ decays.  There are also a number of 
current experiments specializing in charm physics, both in $e^+e^-$ (BES) 
and in fixed-target mode (FOCUS, SELEX, E789, E791).  Results 
from these are beginning to emerge, and the 
next few years should see many interesting developments.
  
\section{$B$ Semileptonic Decays}
\label{sec:BSL}

$B$ physics is all about Standard Model tests and the determination of 
CKM parameters, and semileptonic decays are the core of this program.  
Precise measurements of
$|V_{cb}|$ and $|V_{ub}|$ are the main goals.  Since semileptonic decays
are our main tool, it is essential that we understand this tool very well.  
The last few years
have seen important developments in both theory and experiment.  
We have benefited greatly from the increasingly sophisticated 
application of new theoretical techniques, including Heavy Quark 
Effective Theory (HQET) and lattice gauge calculations.  There has 
been enhanced coordination between experimentalists and theorists, 
and the more recent formation of inter-experiment working groups is 
also proving fruitful.  The challenge has been recognized as having
two essential components: the extraction of all possible information 
from the package of measurements, and consistent and realistic
assessment of theoretical uncertainties.  In this section I address 
three main topics in $B$ semileptonic 
decays.  First I review some long-standing puzzles in the measurements.  
Following that I assess the state of knowledge of $|V_{cb}|$ and of 
$|V_{ub}|$.

\subsection{Puzzles in Semileptonic $B$ Decays}

\begin{figure}[htb!]
\begin{center}
\epsfig{file=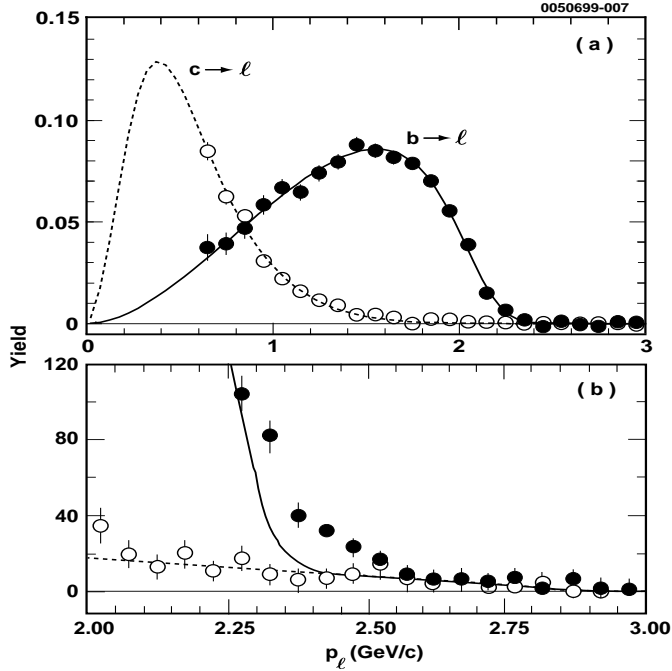,width=3.5in,height=3.5in}
\caption{CLEO's measured lepton spectra: (a) $B \rightarrow X e \nu$
from lepton-tagged analysis; (b) end-point region for 
$B \rightarrow X \ell \nu$ showing excess due to 
$b \rightarrow u \ell \nu$.}
\label{fig:CLEO_lep1}
\end{center}
\end{figure}

Inclusive semileptonic $B$ decay is a beautifully simple process.
Inclusive $b \rightarrow c \ell \nu$ provides the most straightforward 
way to determine $|V_{cb}|$, one which is again acknowledged as
competitive with exclusive determinations.
Inclusive $b \rightarrow u \ell \nu$ gave us the first demonstration
that $V_{ub}$ is nonzero \cite{Fulton:1990pk,Albrecht:1990qv},
and while its interpretation is fraught with
model uncertainties, it remains an important measurement.  
Figure~\ref{fig:CLEO_lep1} is CLEO's snapshot of the entire picture of
semileptonic $B$ decay in the near-threshold environment of the
$\Upsilon(4S)$.  The semileptonic branching fraction and the shape 
of the lepton momentum spectrum are determined using a lepton-tagged
procedure in which charge and angular correlations allow separation
of the primary $b \rightarrow \ell$ and secondary
$b \rightarrow c \rightarrow \ell$ leptons \cite{Barish:1996cx}.
Evidence for charmless decays is revealed as an excess in the region
of the kinematic end point of the $b \rightarrow c \ell \nu$ lepton 
spectrum \cite{Bartelt:1993xh}.

\begin{figure}[htb!]
\begin{center}
\leavevmode
{\epsfxsize=2.5truein \epsfysize=2.5truein \epsfbox{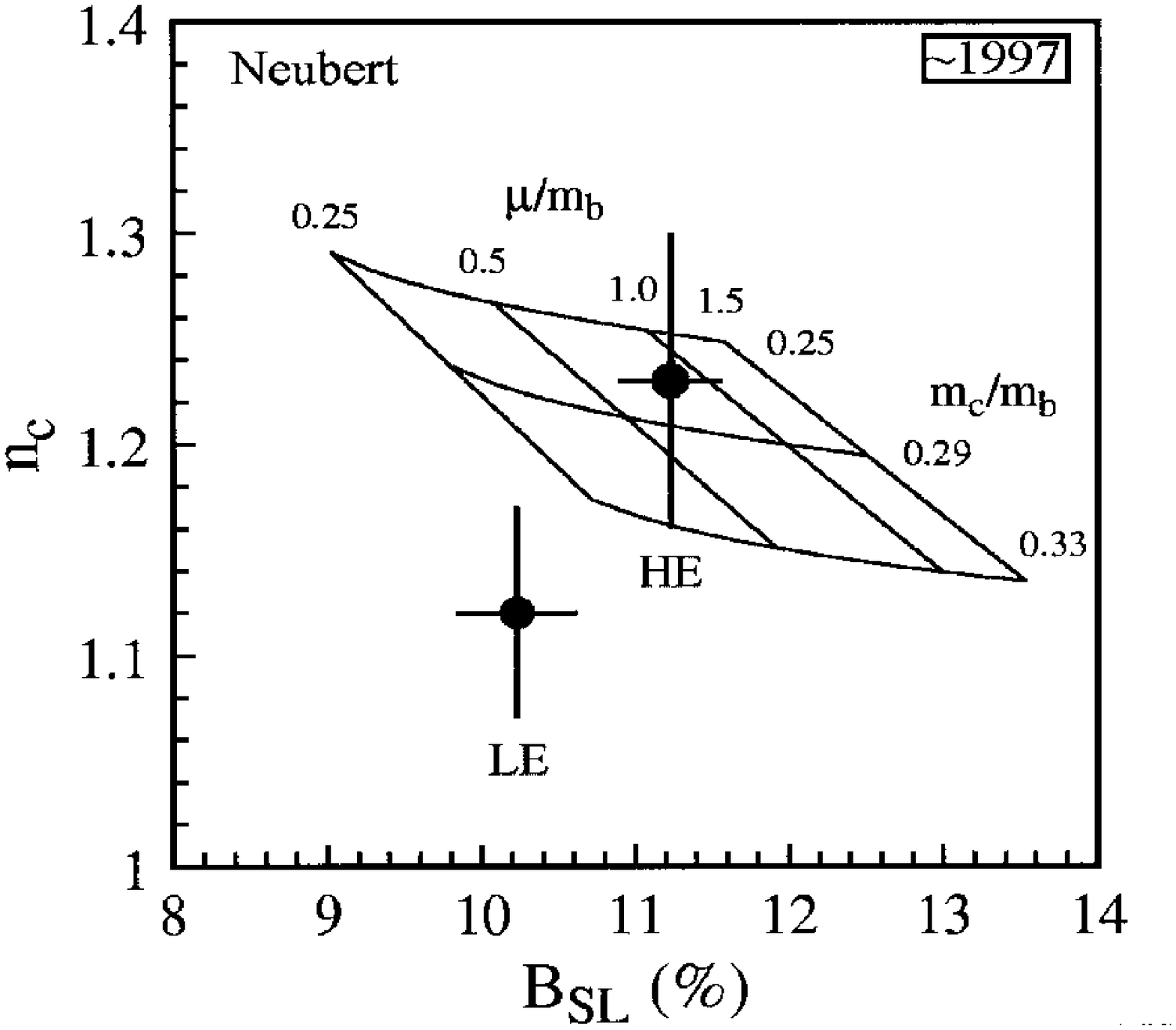}}\hspace*{0.25in}
{\epsfxsize=2.5truein \epsfysize=2.5truein \epsfbox{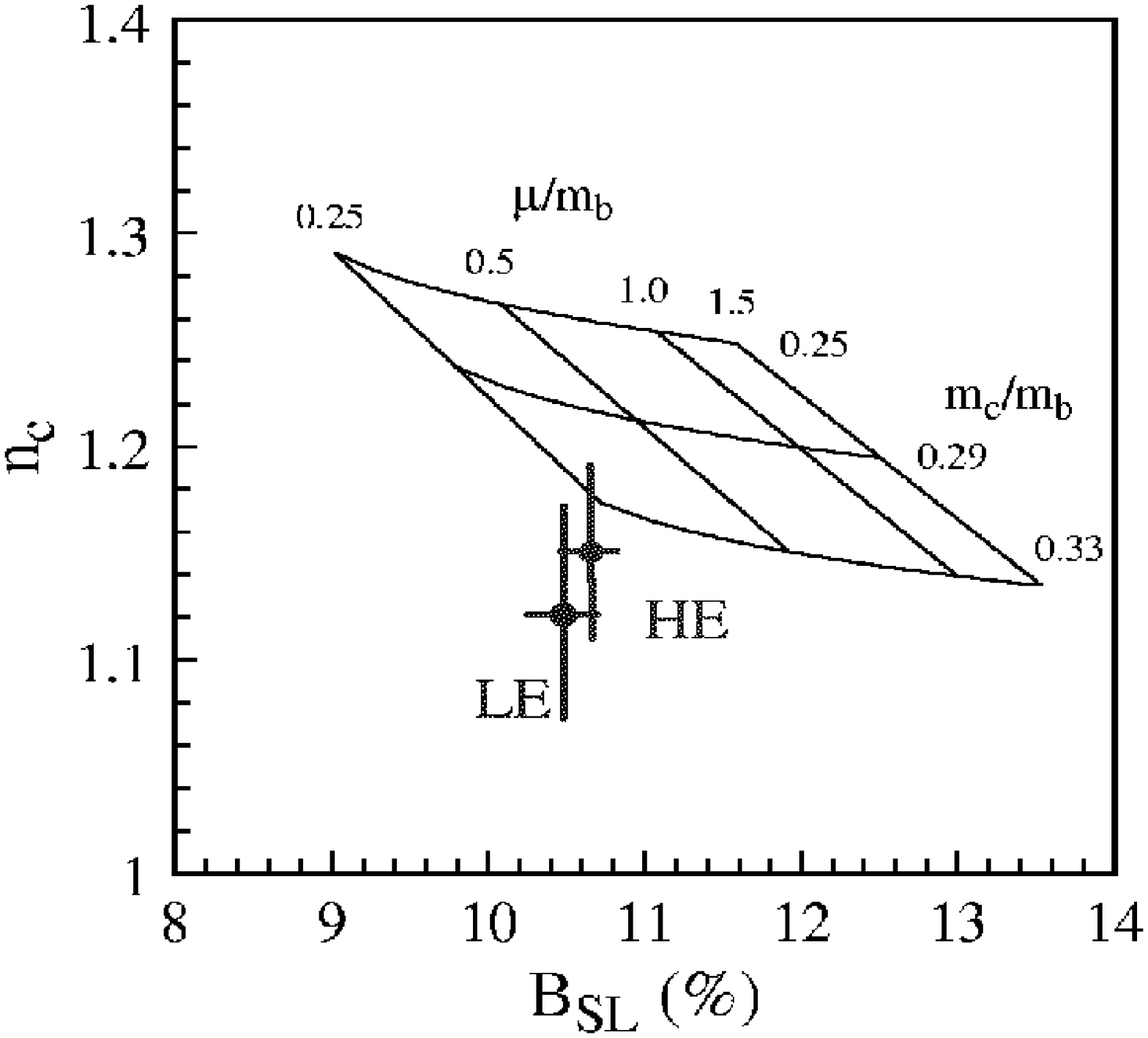}}
\end{center}
\caption{Theory ``comfort zone'' for ${\cal B}(B \rightarrow X \ell \nu)$ 
and charm multiplicity $n_c$.  Parameters are the $m_c/m_b$ quark-mass ratio 
and the normalization scale $\mu$.  
The points on the left (right) are experimental values 
as of 1997 (1999) for experiments at the $Z^0$ (HE) and $\Upsilon(4S)$ (LE).}
\label{fig:Neubert_then_now}
\end{figure}

The simplicity of the semileptonic decay makes it all the more 
vexing that it has been the cause of a great deal of anxiety.  There
are two main puzzles.  Why is the $B$ semileptonic branching 
fraction measured at the $\Upsilon(4S)$ so small?  Why is the $B$ 
semileptonic branching fraction
measured at the $\Upsilon(4S)$ smaller than that at the $Z^0$?
The left-hand graph in Fig.~\ref{fig:Neubert_then_now} shows an assessment
by Neubert of the problem as of about two years 
ago \cite{Neubert:1997gu}.  Naive considerations suggest
a $B$ semileptonic branching fraction of at least 12\%, while experiment
has consistently given values smaller than this.  Mechanisms that 
enhance hadronic $B$ decays can reduce the semileptonic branching 
fraction, but only by increasing $n_c$, the number of charm quarks per $B$.
The data from the $\Upsilon(4S)$ did not bear this out.
The fact that the branching fraction is smaller at the $\Upsilon(4S)$
is a separate matter that is also quite perplexing.  The dominant $B$
mesons at the $Z^0$ are the same as those at the $\Upsilon(4S)$, and the 
inclusion of higher-mass $b$-flavored particles at higher energy 
would be expected to reduce the average semileptonic branching fraction.

No new data from CLEO have been presented since 1997.  There have
been new developments on both the semileptonic branching 
fraction \cite{Gagnon:EPS99} and $n_c$ \cite{Elsing:EPS99} fronts 
from the LEP experiments.  

\begin{figure}[htb!]
\begin{center}
\leavevmode
{\epsfxsize=2.75truein \epsfysize=2.9truein \epsfbox{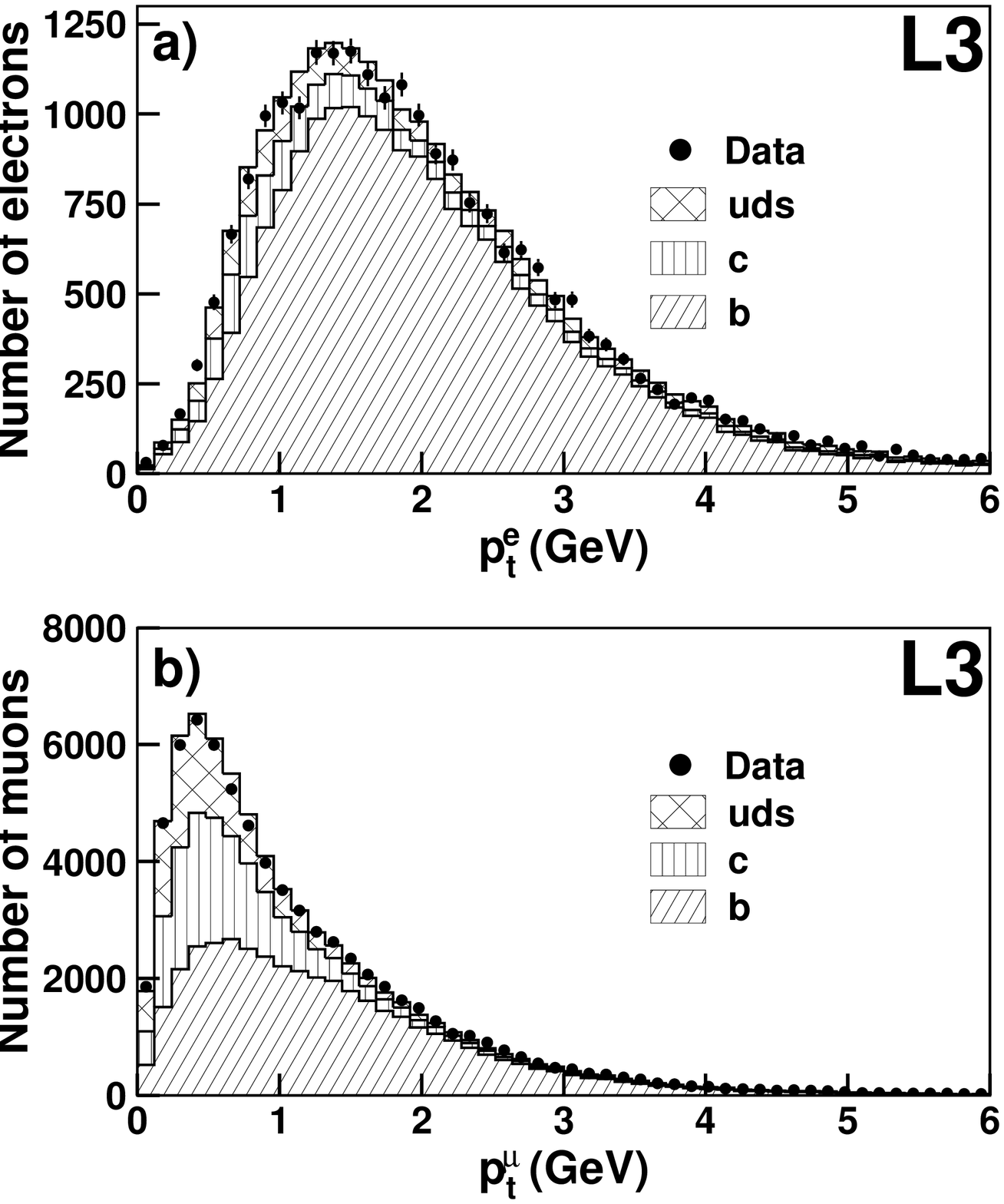}}\hspace*{0.1in}
{\epsfxsize=2.95truein \epsfbox{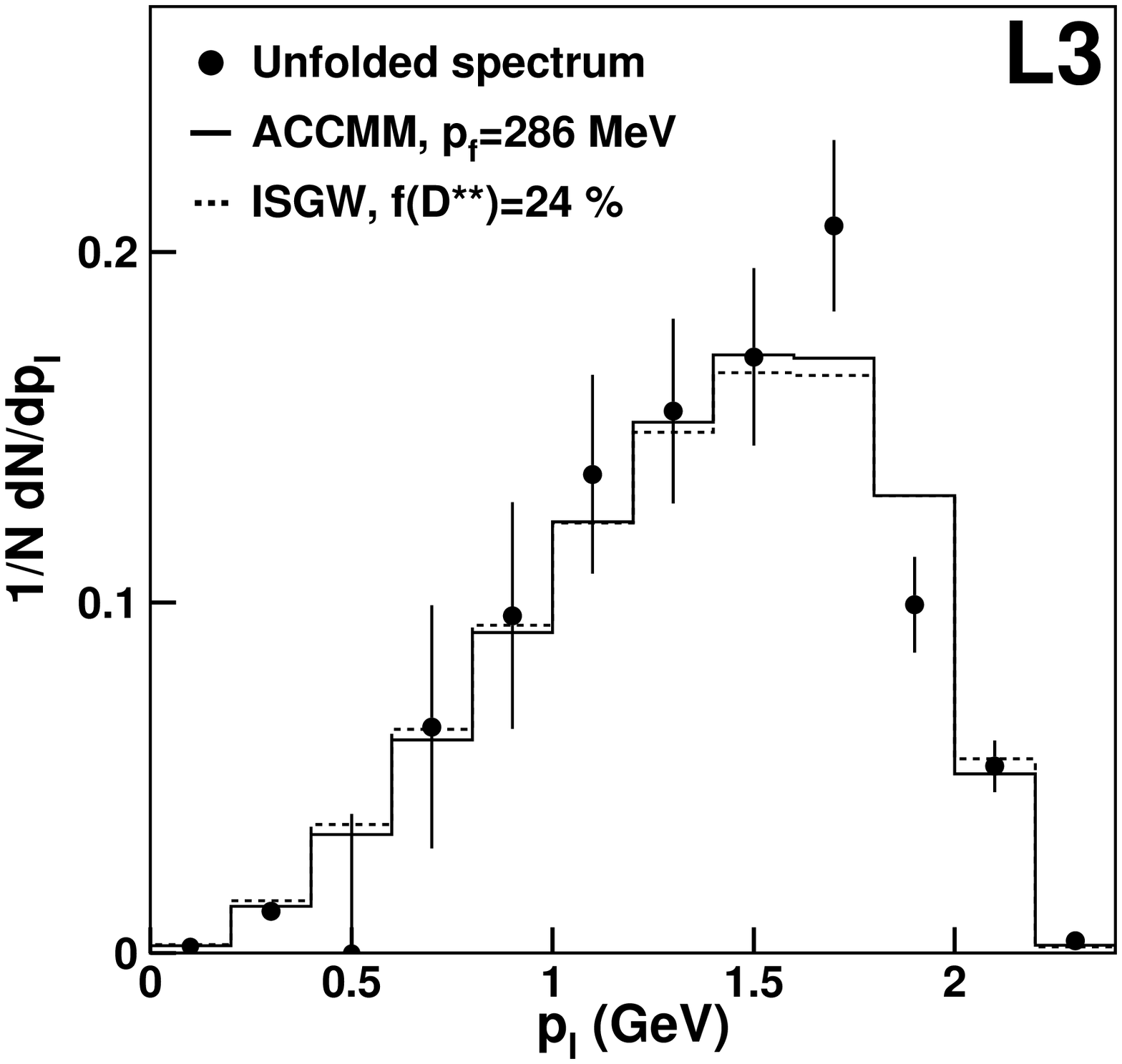}}
\end{center}
\caption{Left: Distributions of transverse momentum with respect to
the closest jet for (a) electrons and (b) muons in L3 data.
Right: Spectrum of lepton momentum in the rest frame of the
decaying $b$ hadron.}
\label{fig:L3SLBR}
\end{figure}

DELPHI (\cite{DELPHISLBR}, L3 \cite{Acciarri:1999ue} and 
OPAL \cite{Abbiendi:1999yy}
have all presented new measurements of the $B$ semileptonic branching
fraction.  They use a variety of techniques with second-lepton,
$B$-vertex and jet-charge tagging, with neural nets employed 
to separate primary, secondary and background leptons.  L3
uses two separate analyses based on double-tag methods to determine 
simultaneously the $Z^0$ $b$-quark fraction $R_b$ and 
${\cal B}(B \rightarrow X \ell \nu$).  One analysis uses a displaced-vertex
$b$ tag, while the other demands a high-$p_t$ lepton.  The observed lepton
$p_t$ distributions and the unfolded momentum in the $b$ rest frame are 
shown in Fig.~\ref{fig:L3SLBR}.  The result of this 
measurement of ${\cal B}(B \rightarrow X \ell \nu$), and of the 
most recent measurements from the other four LEP experiments, are
summarized in Table~\ref{tab:LEPSLBR}.
\begin{table}[htb!]
\begin{center}
\begin {tabular}{l l }
\hline
\hline
Experiment& ${\cal B}(B \rightarrow X \ell \nu)$ (\%) \\ 
\hline
ALEPH 1992-93 data (preliminary 1995)      & $11.01 \pm 0.10 \pm 0.30$ \\
DELPHI 1992-1995 data (preliminary 1999)   & $10.65 \pm 0.07 \pm 0.43$ \\
L3 1994-95 data (preliminary 1999)         & $10.16 \pm 0.13 \pm 0.29$ \\
OPAL 1992-95 data (1999)                   & $10.83 \pm 0.10 \pm 0.26$ \\
\hline
LEP average                                & $10.63 \pm 0.17$ \\
\hline
\end {tabular}
\caption{Recent measurements of ${\cal B}(B \rightarrow X \ell \nu)$
from LEP experiments.}
\label{tab:LEPSLBR}
\end{center}
\end {table}
The new $Z^0$ average, ${\cal B}(B \rightarrow X \ell \nu) 
= (10.63 \pm 0.17)\%$,
represents a significant decrease from the $(11.1 \pm 0.3)\%$ average of
the high-energy measurement circa 1997.  The PDG value for the 
semileptonic branching fraction at the $\Upsilon(4S)$ is 
$(10.45 \pm 0.21)\%$, quite compatible with the new $Z^0$ average.  
This may still overstate the difference, however, since the PDG
average has an aggressively small error considering the 
experimental and theoretical errors of the input
measurements and the spread among them.

New measurements of the multiplicity of charm quarks per $b$ decay
have been reported by ALEPH \cite{ALEPHnc} and 
DELPHI \cite{Abreu:2000vw} .  Combining these with an earlier OPAL 
measurement \cite{Alexander:1996wy} leads to a new
correlated average of 
$n_c = 1.151 \pm 0.022\pm 0.022 \pm 0.051$ \cite{Elsing:EPS99},
where the errors are statistical, systematic, and that due to
input branching fractions.  CLEO's previous number,
$n_c = 1.10 \pm 0.05$ becomes $1.12 \pm 0.05$ when consistent branching
fractions are used, again in very good agreement.

The right-hand graph of Fig.~\ref{fig:Neubert_then_now} is an update of 
Neubert's original comparison.  It is clear from that graph
that the gap between high-energy and low-energy measurements has
narrowed considerably.  The low-energy data still lies outside the
theory comfort zone, but the puzzle seems much less compelling than
it did previously.

\subsection{Determination of $|V_{cb}|$}

We determine the CKM parameter $|V_{cb}|$ by two techniques, both
involving semileptonic decays $b \rightarrow c \ell \nu$.  The 
favored method has been to use the rate for the exclusive
semileptonic decay $B \rightarrow D^* \ell \nu$ (or
$B \rightarrow D \ell \nu$) at zero recoil.  A method that languished
in disrepute for some years, but which has been rehabilitated, is
to use the inclusive semileptonic decay rate.  Both approaches
are rooted in HQET, and there is extensive theoretical guidance on
extracting $|V_{cb}|$ and estimating its uncertainty \cite{Bigi:1999dv}.

The connection between $V_{cb}$ and the semileptonic width $\Gamma_{SL}$
from HQET and the operator product expansion (OPE) is as follows:  
\begin{equation} 
\Gamma _{SL}(B) = \frac{G_F^2m_b^5|V(cb)|^2}{192\pi ^3} 
\left[ z_0\left( 1 - \frac{\mu _{\pi}^2 - \mu _G^2}{2m_b^2}\right) 
- 2 \left( 1 - \frac{m_c^2}{m_b^2}\right) ^4 
\frac{\mu _G^2}{m_b^2} - 
\frac{2\alpha _S}{3\pi }z_0^{(1)} + ... \right] 
\label{eq:MASTER} 
\end{equation} 
Three HQET parameters appear in the expansion.  $\bar \Lambda$ connects the 
quark mass with the meson mass.  $\mu_\pi^2$ (or its relative $\lambda_1$) 
relates to the average kinetic energy of the $b$ quark.  $\mu_G^2$ 
($\lambda_2$) is connected to the hyperfine splitting.  Bigi judges that 
a ``prudent'' theoretical uncertainty for the extraction of $|V_{cb}|$ by 
this procedure is $\sim 6\%$ \cite{Bigi:1999dv}.  The contributions of the 
uncertainties in the experimental inputs, the $B$ semileptonic branching 
fraction ($(10.5 \pm 0.2 \pm 0.4)\%$) and average $B$ lifetime 
($1.61 \pm 0.02$~ps), are small in comparison.  The result is  
$|V_{cb}| = (40. \pm 0.4 \pm 2.4) \times 10^{-3}$.

On the exclusive front, HQ symmetry tells us that a heavy-light meson
decaying at rest really isn't changing at all.  A measurement of the
decay rate of $B \rightarrow D^* \ell^- {\bar \nu}$ at maximum $q^2$ 
($w=1$) gives ${\cal F}(1)V_{cb}$:
\begin{eqnarray} 
\frac{d \Gamma}{d w} & = & \frac{G_F^2}{48\pi^3} {|V(cb)|^2} m_{D^*}
(m_B - m_{D^*})^2 {\cal F}^2(w) {\cal G}(w)\mbox{, where} \\
w & = & (m_B^2 + m_{D^*}^2 - q^2)/(2 m_B m_{D^*}) \nonumber
\label{eq:VcbDstar} 
\end{eqnarray} 
Add the form-factor normalization ${\cal F}(1)$ from theory and we are 
done.  There has been
continuing evolution in thinking about ${\cal F}(1)$, and some 
controversy \cite{VcbFs}.  Bigi \cite{Bigi:1999dv} suggests
${\cal F}(1) = 0.88 \pm 0.08$, with a smaller value and a much bigger
error than earlier suggestions.

\begin{figure}[htb!]
\begin{center}
\leavevmode
{\epsfxsize=2.95truein \epsfysize=3.1truein \epsfbox{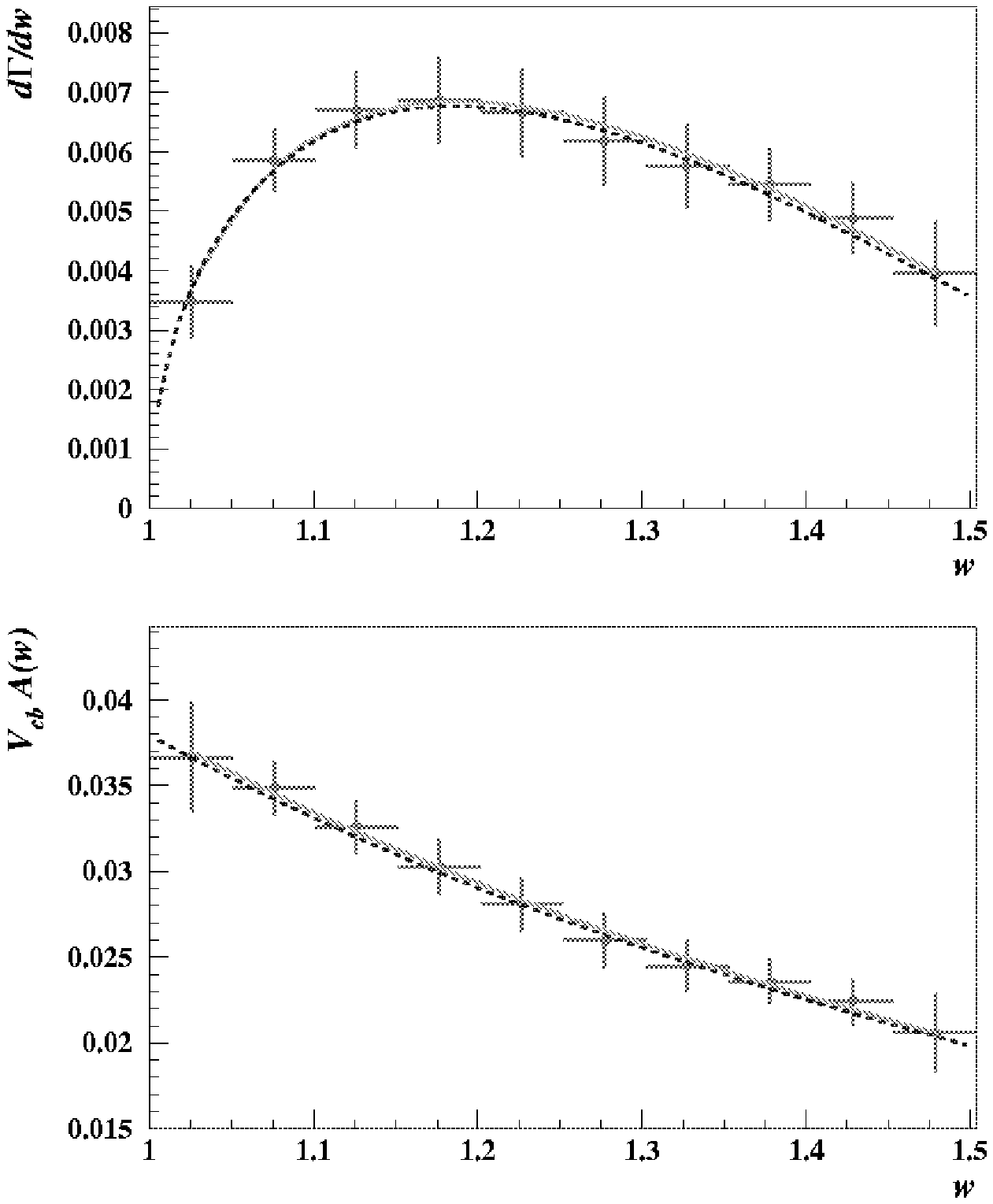}}\hspace*{0.01in}
{\epsfxsize=3.truein \epsfysize=3.truein \epsfbox{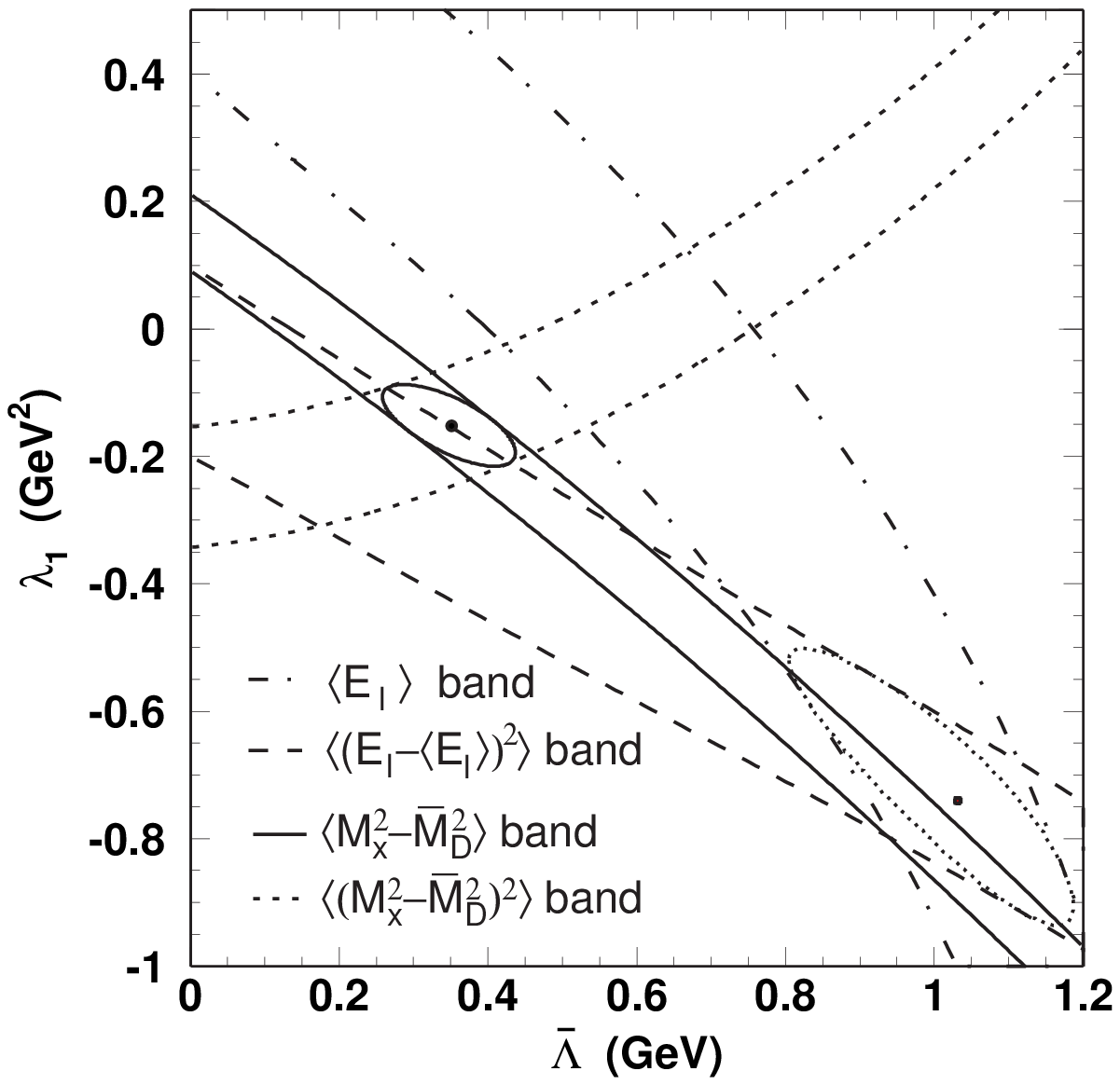}}
\end{center}
\caption{Left: Preliminary DELPHI results on 
$B \rightarrow D^* \ell^- {\bar \nu}$.  The upper graph is the unfolded
distribution of differential decay width, and the lower graph shows 
the extraction of ${\cal F}(1) |V_{cb}|$.
Right: Bands in HQET/OPE parameters $\lambda_1$ and $\Lambda$
found in CLEO's preliminary analysis of the first two moments of the
$B$ semileptonic decay recoil-mass-squared and lepton-energy
moments.  The 1$\sigma$ error ellipses are shown.}
\label{fig:DELPHI_and_moments}
\end{figure}

A new measurement of $B \rightarrow D^* \ell^- {\bar \nu}$ has been 
reported by DELPHI \cite{DELPHIBDs}, joining ALEPH \cite{Buskulic:1997yq}, 
OPAL \cite{Ackerstaff:1997dc}, and CLEO \cite{Barish:1995mu}.  The new 
DELPHI measurement (Fig.~\ref{fig:DELPHI_and_moments}) is based on 
$\sim 5500$ tagged decays and has the best precision.
(CLEO has so far reported on only one sixth of its total data sample.)  
Table~\ref{tab:exclVcb} summarizes the results on $|V_{cb}|$
from $B \rightarrow D^* \ell^- {\bar \nu}$, following 
the LEP $V_{cb}$ working group \cite{Calvi:HF8}, and Bigi's
\begin{table}[tbh!]
\vspace{-2mm}
\begin{center}
\caption{Determinations of $|V_{cb}|$ using 
$B\to D^*\ell^-\overline{\nu}$ decays
at $\omega = 1$. 
\label{tab:exclVcb}}
\vskip 0.1 in
\begin{tabular}{|l|c|}\hline
Experiment & $V_{cb}$ $(\times 10^{-3})$\\
\hline

\hline
ALEPH \cite{Buskulic:1997yq} & $36.6\pm 2.4 \pm 1.8$ \\
DELPHI \cite{DELPHIBDs} & $41.2\pm 1.6 \pm 2.8$ \\
OPAL \cite{Ackerstaff:1997dc} & $38.9\pm 2.2 \pm 3.1$ \\
\hline
LEP weighted average & $38.4 \pm 1.1 \pm 2.2 \pm 2.2$ \\
\hline
CLEO\cite{Barish:1995mu} & $39.4\pm 2.1 \pm 2.0 \pm 1.4$ \\
\hline
\end{tabular}
\end{center}
\end{table}
proposal for ${\cal F}(1)$.
Everything agrees very well.  The exclusive $|V_{cb}|$ result is consistent 
with the inclusive, and the overall precision is comparable.  

Both
extraction procedures rely on the HQET/OPE approach, which is beautiful 
but largely unvalidated by experiment.  Experimental tests are needed,
and measurements of the parameters ${\bar \Lambda}$ and 
$\lambda_1$/$\mu_\pi^2$ would be extremely
valuable.  Measurements of the moments of the hadronic mass and lepton
energy in $B$ decays have been proposed to do 
this \cite{Voloshin:1995cy,Falk:1996kn,Falk:1996me,Gremm:1997df}.  
CLEO has made a preliminary measurement of this
type \cite{CLEO:moments}, the results of which are shown in
in Fig.~\ref{fig:DELPHI_and_moments}.
The mass moments and lepton-energy moments do not admit a common
solution, and the discrepancy is significant.  Perhaps one (or
both) of the measurements is flawed, or perhaps there is something
wrong with the theoretical approach.  Some have suggested that the 
assumption of quark/hadron duality should be scrutinized.  CLEO is 
updating its measurements with more data and a better understanding
of the experimental systematics.

\subsection{Determination of $|V_{ub}|$}

Compared to $|V_{ub}|$, $|V_{cb}|$ was easy.  The advantages afforded by
heavy-quark symmetry in studying $b \rightarrow c \ell \nu$ do not carry
over to the heavy-to-light transition of $b \rightarrow u \ell \nu$.  
Extraction of $|V_{ub}|$ is highly model-dependent, the experiments
are tougher, and the achievable precision will likely always be less.
The CLEO and ARGUS discovery measurements for 
$b \rightarrow u \ell \nu$ \cite{Fulton:1990pk,Albrecht:1990qv},
and the subsequent confirmation in CLEO~II data \cite{Bartelt:1993xh}
were based on the nonzero excess of leptons near and above the
kinematic limit for $b \rightarrow c \ell \nu$ at
the $\Upsilon(4S)$.  The measurement of the yield is
straightforward, but since only a tiny corner of the 
$b \rightarrow u \ell \nu$ phase space is sampled, 
models \cite{isgw,isgw2,vubmods} must be used to extrapolate to 
the total rate.  It is very difficult to assess the theoretical 
uncertainty, and my preference is to be very cautious:
$|V_{ub}/V_{cb}| = 0.08 \pm 0.02$.

\begin{figure}[htb!]
\begin{center}
\leavevmode
{\epsfxsize=2.5truein \epsfysize=2.5truein \epsfbox{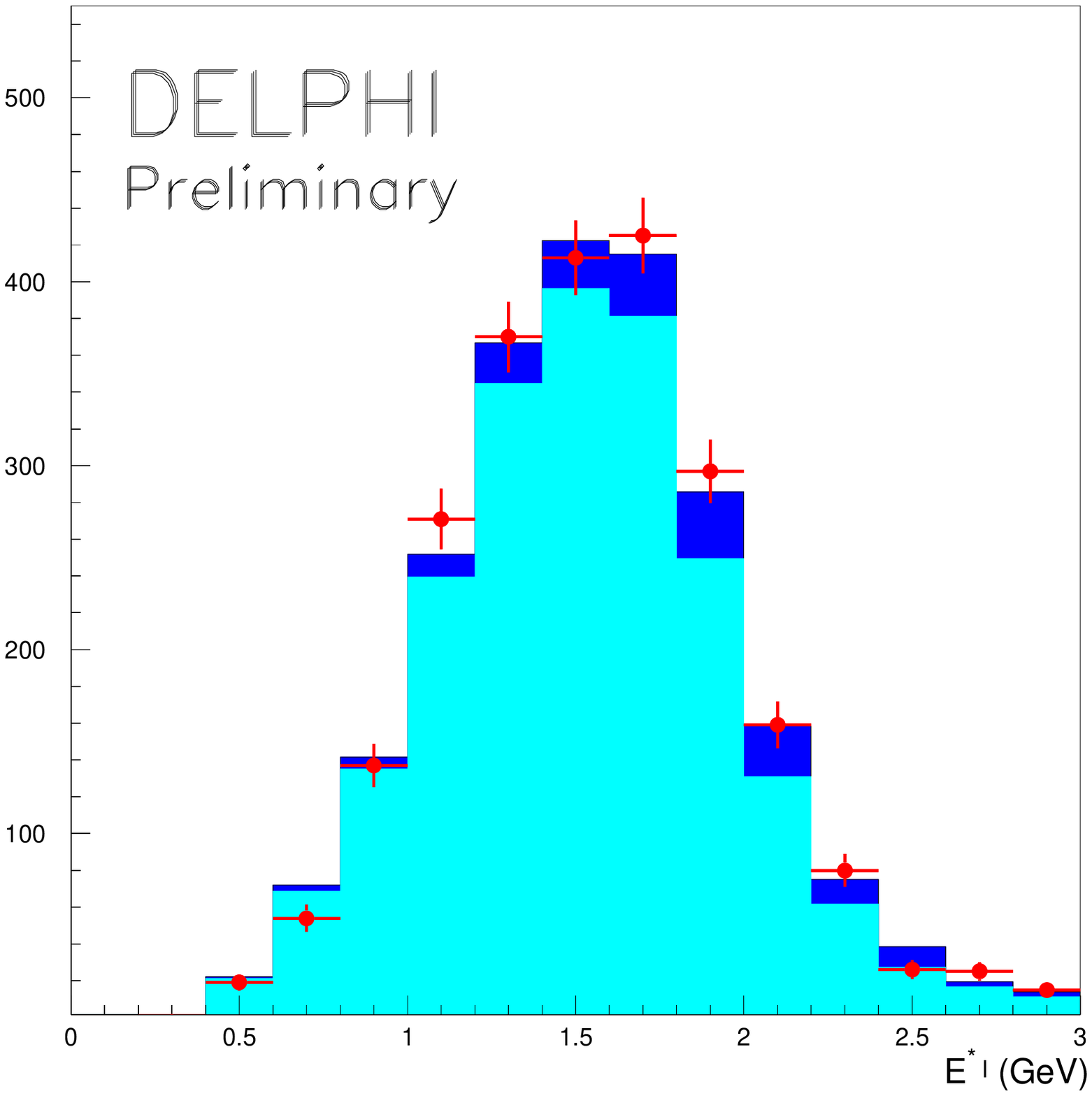}}\hspace*{0.15in}
{\epsfxsize=2.5truein \epsfysize=2.5truein \epsfbox{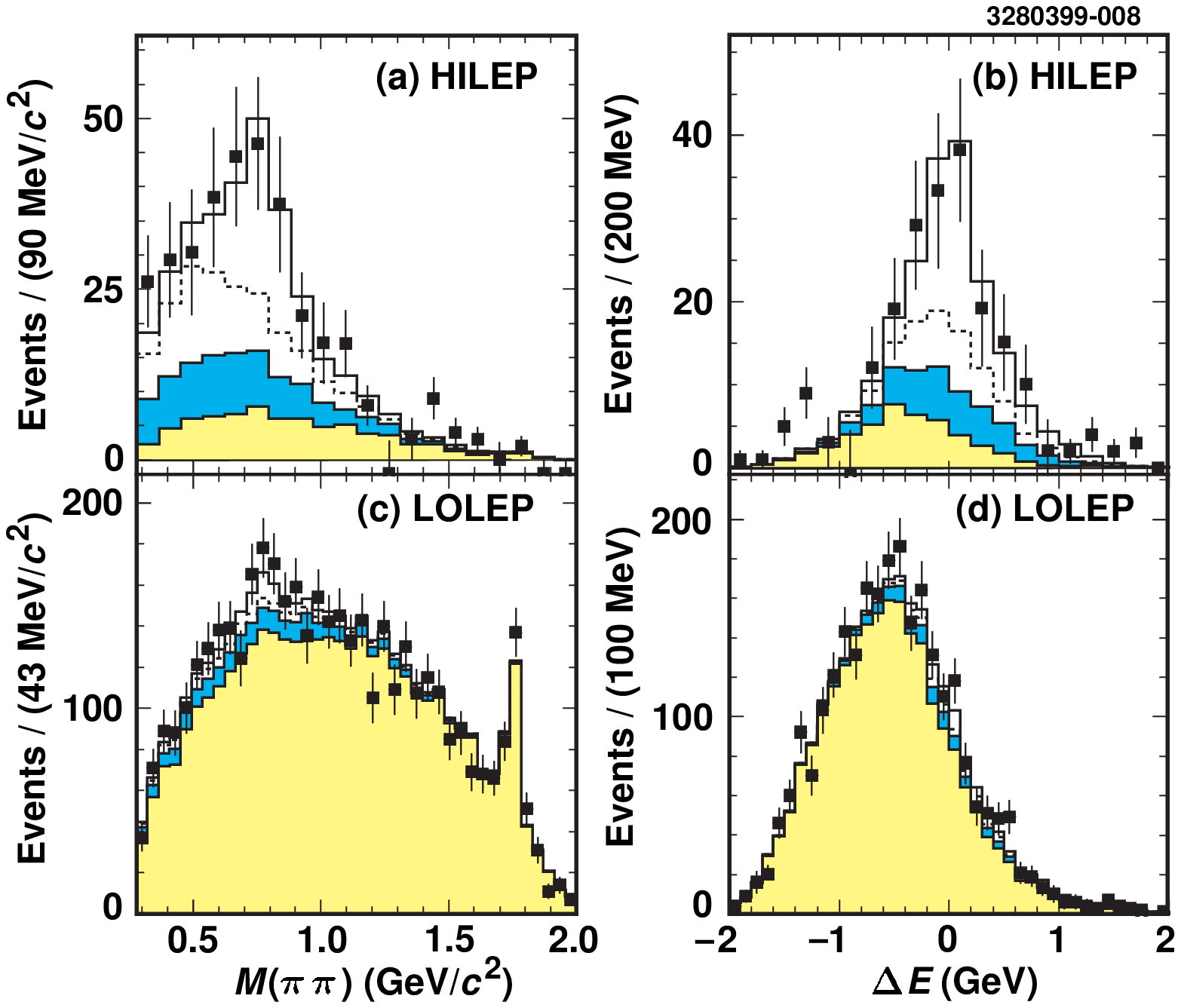}}
\end{center}
\caption{Left: The lepton energy in the $B$-meson rest frame for
the DELPHI $b \rightarrow u$-enriched sample.  The light-shaded histogram 
is the fitted background and the dark-shaded histogram is the fitted 
signal.  Right: Fit projections for CLEO's $B \rightarrow \rho \ell \nu$ 
analysis showing signal and background components in two 
momentum ranges (HILEP: $p>2.3 {\rm GeV}$, LOLEP: $2.0<p<2.3 {\rm GeV}/c$.}
\label{fig:DELPHI_CLEO_bu}
\end{figure}

In the past few years, ALEPH \cite{Barate:1999vv}, 
L3 \cite{Acciarri:1998if} and 
DELPHI \cite{DELPHIvub} have all presented 
ambitious analyses that seek to measure the $b \rightarrow u \ell \nu$
component in $b$ decays at the $Z^0$.  The strategy is to reconstruct
the charmless hadronic mass $m_X$ in $b \rightarrow X \ell \nu$, and
to enrich the sample in $b \rightarrow u \ell \nu$ by demanding
$m_X$ to be less than $\sim 1.6~{\rm GeV}/c^2$.  Discrimination between 
$b \rightarrow u$-like and $b \rightarrow c$-like
is based on many event details, including displaced vertices, transverse
momentum, presence of kaons, and other features, combined for maximum
discrimination with neural nets.  This technique exploits the advantages
of production at the $Z^0$: well-separated jets and fast-moving $B$'s,
but it requires very detailed understanding of $b \rightarrow c \ell \nu$.
The DELPHI $b \rightarrow u \ell \nu$ lepton-energy distribution in 
the $B$ rest frame is shown in Fig.~\ref{fig:DELPHI_CLEO_bu}.  It is fitted to 
signal and background components to extract $|V_{ub}/V_{cb}|$.
The LEP $|V_{ub}|$ working group \cite{Calvi:HF8} combines the three
LEP measurements to obtain an average of
$|V_{ub}|=(4.05^{+0.62}_{-0.74}) \times 10^{-3}$, very consistent
with CLEO.  Because more
of the spectrum is measured than in the end-point analysis, the
extraction of $|V_{ub}|$ should have less theoretical uncertainty in 
principle.  Unfortunately, dealing with the enormous 
$b \rightarrow c \ell \nu$ component introduces different uncertainties 
that are also very difficult to quantify.  

The first measurement of the exclusive charmless semileptonic decays
$B \rightarrow \pi/\rho \ell \nu$ by CLEO \cite{Alexander:1996qu} was a
milestone in the determination of $|V_{ub}|$.  Conventional wisdom has 
held that the extraction of $|V_{ub}|$ from exclusive decays would
be less model-dependent than the earlier end-point measurements.
The main reason for this prejudice has been that tools like light-cone
sum rules and lattice QCD, along with experimental input from charm
decays, would provide necessary form-factor information.

CLEO has presented a new analysis of 
$B \rightarrow \rho \ell \nu$ \cite{Behrens:2000vv}
with higher efficiency than full reconstruction. 
Binned maximum-likelihood fits
are made of the lepton energy, $\Delta E$ and candidate mass to 
parameterizations for 
$B \rightarrow \rho \ell \nu$, $B \rightarrow \pi \ell \nu$,
$B \rightarrow \omega \ell \nu$, other $b \rightarrow u \ell \nu$,
continuum, and fake leptons.  
The data sample is divided according to lepton-momentum, with
the greatest sensitivity to $B \to \rho \ell \nu$ in the 
highest bin ($>2.3 {\rm GeV}/c$).  Several models \cite{isgw2,rholnumods} 
are used to evaluate efficiencies and extract $|V_{ub}|$.
This measurement is averaged with the previously published CLEO
results \cite{Alexander:1996qu}, to give
${\cal B}(B^0 \rightarrow \rho^- \ell^+ \nu) = (2.57 \pm 0.29^{+0.33}_{-0.46}
\pm 0.41) \times 10^{-4}$ and
$\vert V_{ub} \vert = (3.25 \pm 0.14 ^{+0.21}_{-0.29} \pm 0.55) 
\times 10^{-3}$.  The experimental uncertainties in this measurement
are smaller than the theoretical uncertainty.  Further progress
will depend on advances in theory with guidance from experiment.
This analysis included a first measurement of the $q^2$ distribution
for $B \rightarrow \rho \ell \nu$, but the data are not yet sufficiently 
precise to discriminate among models.

\section{Rare $B$ Decays}
\label{sec:RareB}

Rare decays have provided much of the excitement in $B$ physics
during the past several years.  As data samples have grown, the roster of
rare processes that have come within the reach of experiment has
lengthened steadily.  The discovery of the electroweak penguin decay
$b \rightarrow s \gamma$, first exclusively \cite{Ammar:1993sh} and
later inclusively \cite{Alam:1995aw}, was a
major milestone in two ways.  First, it excluded a broad range
of physics beyond the Standard Model by coming in very close to
expectations \cite{Hewett:ay}.  Second, it was a first signal of the 
major role of penguin processes in $B$ decays, a feature that has 
greatly influenced expectations for studies of CP violation.

The principal source for new measurements of rare $b$ decays has been
the nearly 20 million $B$ mesons in the CLEO~II/II.V data sample.
Contributions have also been made by ALEPH,
and by CDF and D0 in searches for modes with dileptons.  

\subsection{Charmless Two-Body $B$ Decays}

\begin{figure}[tb!]
\begin{center}
\epsfig{file=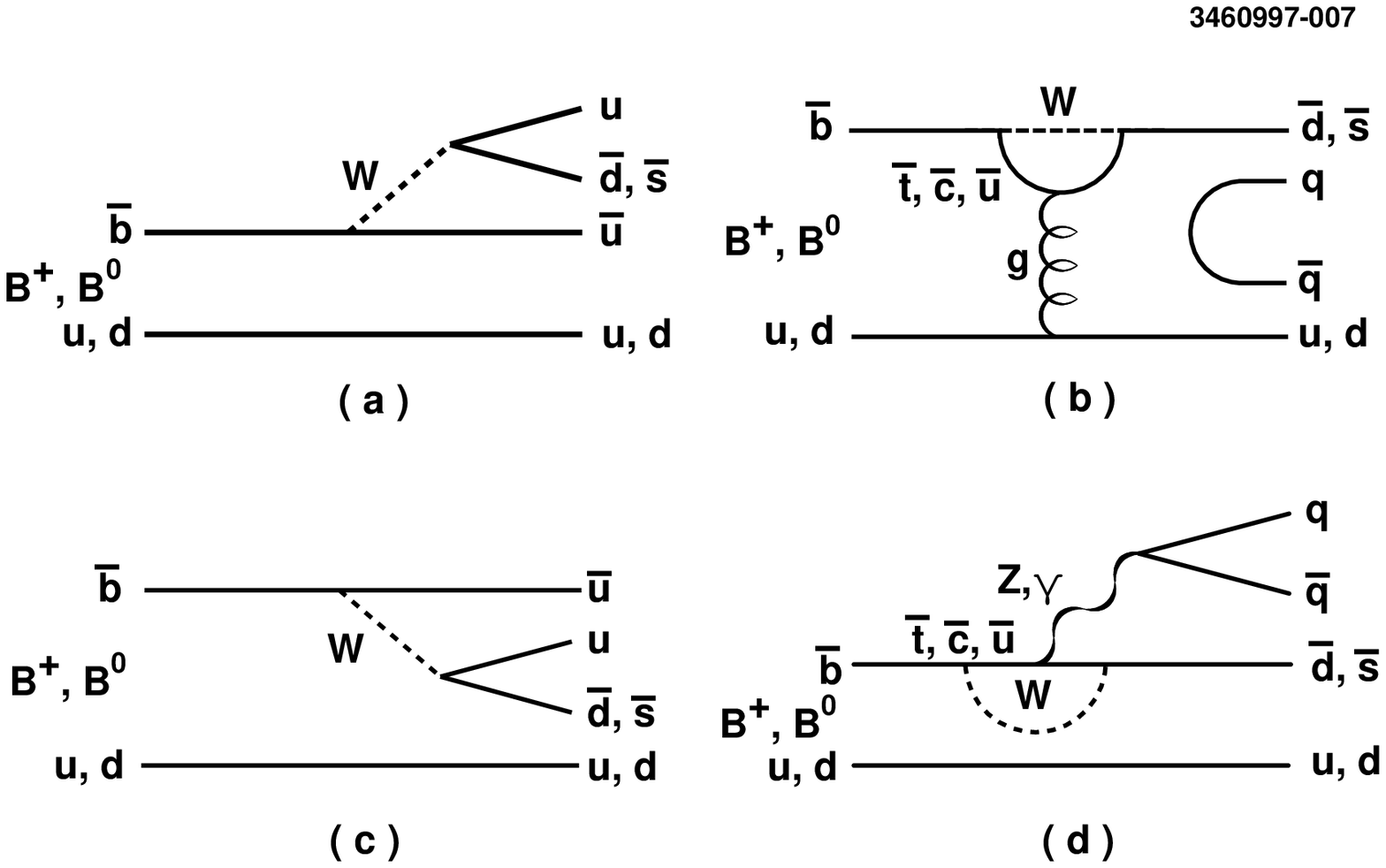,height=2.5in}
\caption{Principal contributors to charmless $B$ decays:
(a) external $W$-emission, (b) gluonic penguin, (c) internal $W$-emission,
(d) external electroweak penguin.}
\label{fig:rare_feyn}
\end{center}
\end{figure}

CLEO has made great strides in filling in the table of charmless 
two-body decays.  The implications of these measurements for the
future $B$ program are significant.  The principal contributing 
processes, $b \rightarrow s$ penguins and $b \rightarrow u$ trees,
are shown in Fig.~\ref{fig:rare_feyn}.  
Interference between tree and penguin diagrams opens a window on the 
unitarity-triangle angle $\gamma$ in measurements of decay rates. 
CLEO measurements of $B \rightarrow \pi \pi$, $B \rightarrow \pi \rho$ 
and other modes define the strategies for future CP-violation
searches, including the determination of $\alpha$.  Searches for direct 
CP violation could provide our first glimpse of physics beyond the 
Standard Model.

CLEO's two-body charmless decay analyses share
a common set of tools that take advantage of the features of
$B {\bar B}$ production at the $\Upsilon(4S)$.  Candidates are identified
based on the beam-constrained mass, $M_B = \sqrt{E^2_{beam} - |{\vec p}|^2}$,
and the difference between the beam energy and the measured energy of the 
$B$ candidate's daughters, $\Delta E = E_1 +E_2 - E_{beam}$.  For signal events
$M_B$ must be close to the known $B$-meson mass 
($\sigma(M_B) \simeq 2.5$~MeV), and $\Delta E$ must be close to zero
($\sigma(\Delta E) \simeq 15-25$~Mev, depending on the mode).  
Two-body $B$ decays have considerable background
from continuum $e^+e^- \rightarrow q {\bar q}$, for which the cross section
is roughly three times higher than $B {\bar B}$.  The jet-like continuum
background is aggressively suppressed with event-shape cuts based on
numerous input variables that are combined into a linear multivariate
(Fisher) discriminant.  Residual continuum background is estimated with data
collected 60~MeV below the $\Upsilon(4S)$ resonance.  In 
addition to the common selection criteria, there are a number of 
signal-specific cuts including resonance mass, particle identification 
and helicity angles.

After imposition of loose cuts, final signals are extracted 
with unbinned maximum likelihood fits to $\sim 7$ quantities, including 
$M_B$, $\Delta E$, resonant masses ($\rho$, $K^*$, $\eta$, $\eta'$,
$\omega$), particle ID, helicity angles, and continuum-suppression 
variables.  In addition to the fits, cut-and-count analyses are
performed for confirmation.  All of CLEO's new preliminary results have 
been obtained using between 5.8 million and the full 9.7 million 
$B {\bar B}$ events in the combined CLEO~II and CLEO~II.V data sets.

\begin{figure}[htb!]
\begin{center}
\leavevmode
{\epsfxsize=2.65truein \epsfbox{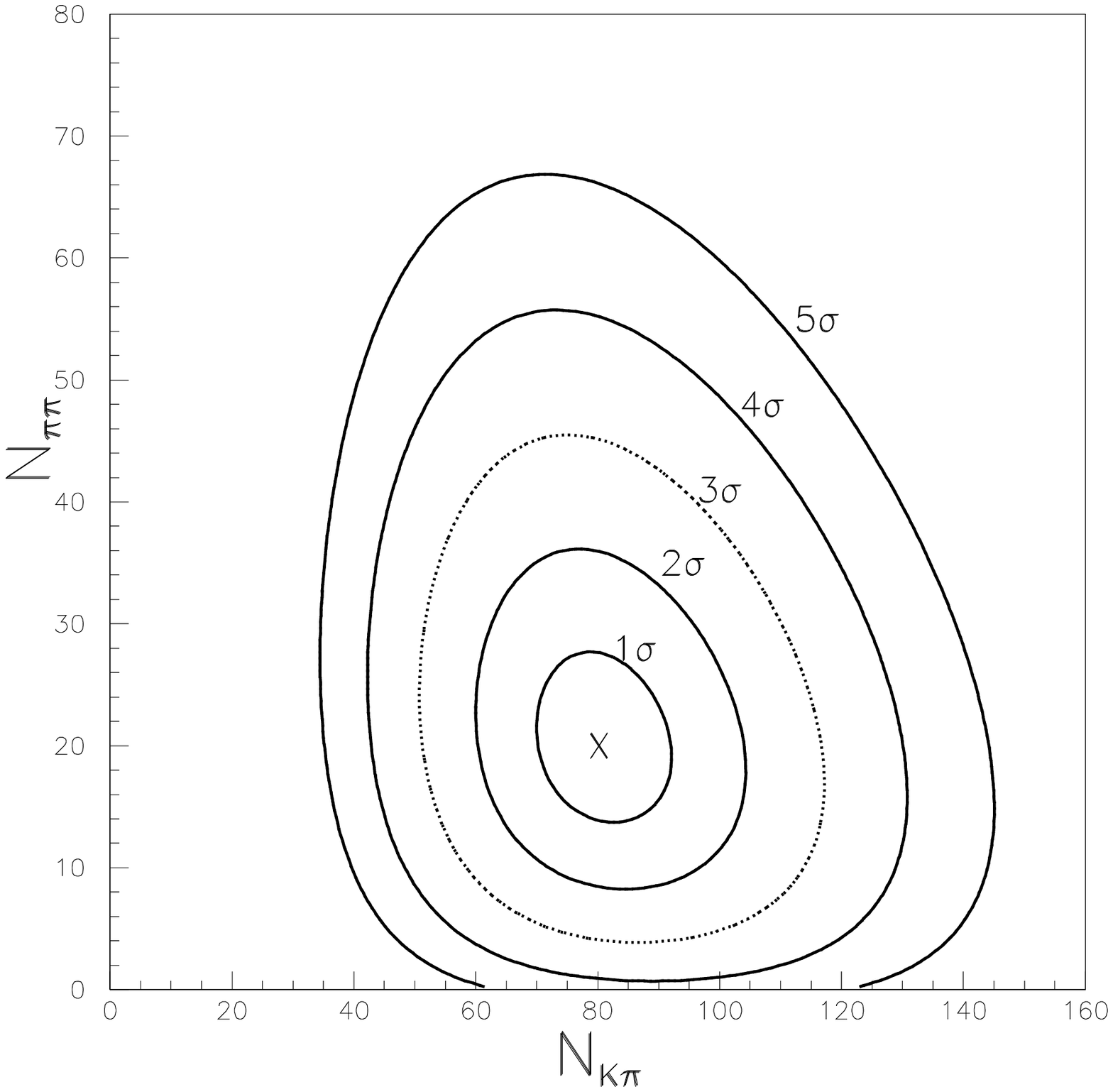}}\hspace*{0.05in}
{\epsfxsize=2.75truein \epsfbox{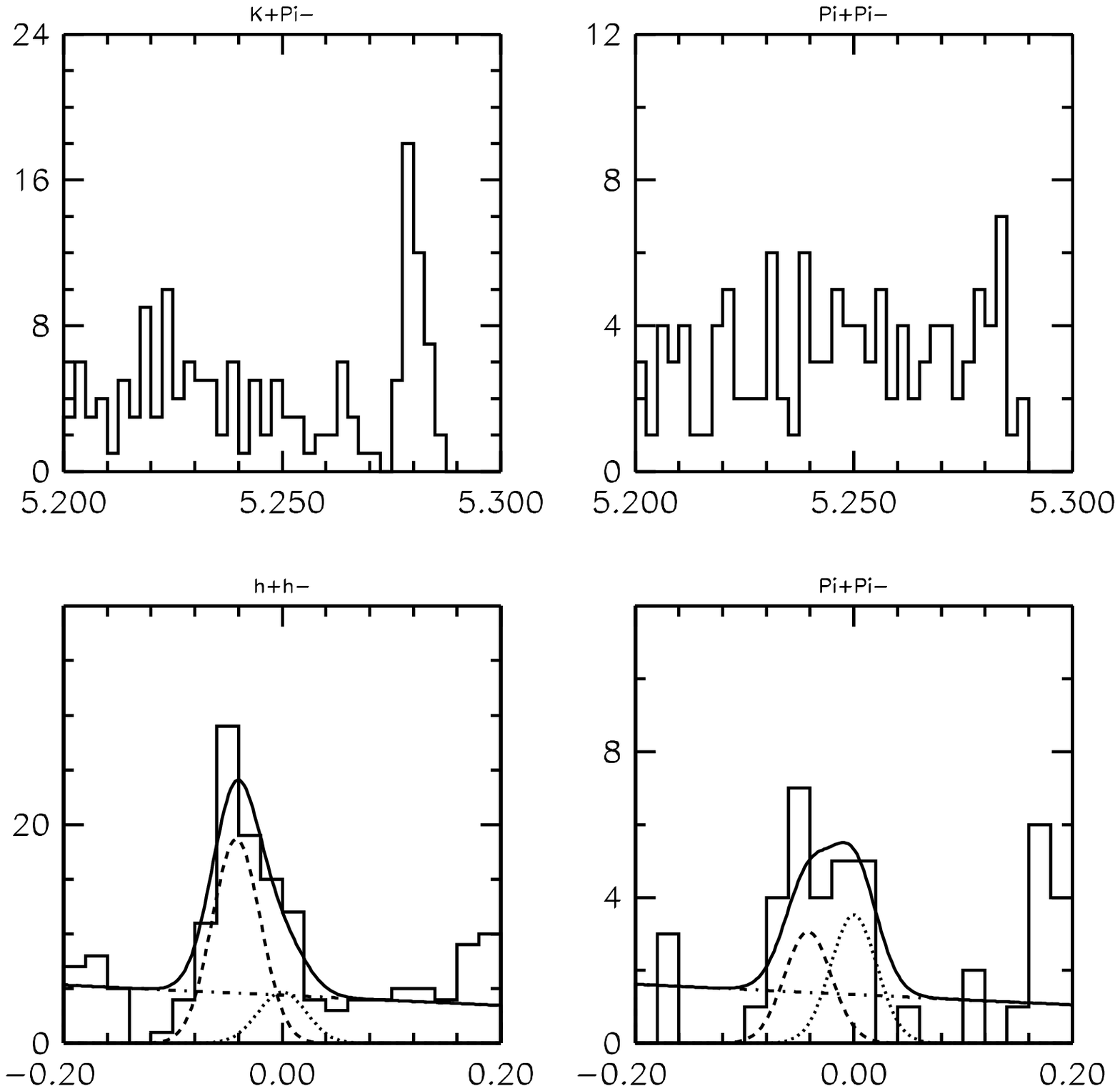}}
\end{center}
\caption{Left: Likelihood contours (statistical errors only) for the
maximum likelihood fit yields for $B \rightarrow \pi^+ \pi^-$ vs. 
$B \rightarrow K^\pm \pi^\mp$.  Right: Projections of $K \pi$ and
$\pi \pi$ events onto $M_B$ and $\Delta E$.}
\label{fig:CLEO_pipi_fig1}
\end{figure}

Results from CLEO's updated search for the decays 
$B \rightarrow h^+ h^-$ \cite{Kwon:1999hx} are shown in 
Fig.~\ref{fig:CLEO_pipi_fig1}.    
The full likelihood fit of 
$B \rightarrow \pi^+ \pi^-$ and $B \rightarrow K^\pm \pi^\mp$ gives 
the confidence-level contours shown, with 11.7$\sigma$ and 4.2$\sigma$
statistical significance, respectively. 
(There is no evidence for $B \rightarrow K^+ K^-$.)
The projected signal for the $K^\pm \pi^\mp$
mode is extremely strong, while that for $\pi^+ \pi^-$ is much less
compelling.  Projections onto $\Delta E$ for the undifferentiated
$h^+ h^-$ events analyzed as $\pi^+ \pi^-$ again show the dominance
of $K^\pm \pi^\mp$, with no clear indication of a $\pi^+ \pi^-$ signal.
For the events satisfying particle-ID criteria for $\pi^+ \pi^-$,
however, it is clear that a $\pi^+ \pi^-$ component is needed in addition
to the misidentified $K^\pm \pi^\mp$ to explain the distribution.
The results for the branching fractions of these modes are given in
Table~\ref{tab:CLEO_Kpi_results}.  
\begin{table}[htb!]
\begin{center}
\begin {tabular}{l c c c c }
\hline
\hline
Mode& ${\cal E}(\%)$ & ${\cal B}_{fit}$($10^{-6}$) & Signif.(std. dev.) 
&${\cal B}$($10^{-6}$) \\
\hline
$\pi^+\pi^-$ &
$45$
& $4.7^{+1.8}_{-1.5}$  & 4.2  
& $4.7^{+1.8}_{-1.5}\pm 0.6$       \\
$\pi^+\pi^0$ & 
$41$
& $5.4^{+2.1}_{-2.0}\pm 1.5$ & 3.2 
&$<12$         \\
\hline
$K^+\pi^-$   &
$45$
& $18.8^{+2.8}_{-2.6}$ & 11.7
& $18.8^{+2.8}_{-2.6} \pm 1.3$        \\
$K^+\pi^0$   &
$38$
& $12.1^{+3.0}_{-2.8}$  & 6.1
& $12.1^{+3.0+2.1}_{-2.8-1.4}$         \\
$K^0\pi^+$   
& $14$
&$18.2^{+4.6}_{-4.0}$ & 7.6
&$18.2^{+4.6}_{-4.0} \pm 1.6$               \\
$K^0\pi^0$   
& $11$
&$14.8^{+5.9}_{-5.1}$ & 4.7
&$14.8^{+5.9+2.4}_{-5.1-3.3}$               \\
\hline
$K^+K^-$     
& $45$
&  & 0.
& $<2.0$  \\
$K^+\bar{K}^0$   
& $14$
& $$  & 1.1 
& $<5.1$    \\
\hline
\hline
\end {tabular}
\caption{Results from CLEO's $B \rightarrow h^+ h^-$ analysis.  
The reconstruction efficiency 
${\cal E}$ includes branching fractions for 
$K^0 \to K^0_S \to \pi^+\pi^-$ and $\pi^0\to \gamma\gamma$. We quote 
the central-value branching fraction in $\pi^\pm\pi^0$ for convenience 
only; the statistical significance 
for this mode is insufficient for a first observation.}
\label{tab:CLEO_Kpi_results}
\end{center}
\end {table}

This first measurement of 
$B \rightarrow \pi^+ \pi^-$ provides a long-awaited
piece of the rare-decay puzzle, and it confirms that studies of this 
mode, and its future use in CP-violation measurements, are greatly 
complicated by ``penguin pollution.''  This study is 
only one piece of a growing picture, however, and CLEO has also 
presented new results on the closely related decay modes 
$B^+ \rightarrow K^0 h^+$ and $B^+ \rightarrow h^+ \pi^0$,
also summarized in Table~\ref{tab:CLEO_Kpi_results} 
\cite{Kwon:1999hx}.
In this case there are statistically significant signals for
$K^0 \pi^+$ ($7.6 \sigma$) and $K^+ \pi^0$ ($6.1 \sigma$), but not
for $\pi^+ \pi^0$, reinforcing the picture of penguin dominance.
In addition, there has been a first neasurement of the
decay to $K^0 \pi^0$, providing a complete set of four $K \pi$ branching
fractions.

As the available sample of charmless hadronic $B$ decays grows, it
becomes possible to search for direct CP violation.
CP asymmetries are possible when two or more contributing diagrams
differ in weak and strong phases.  CLEO has presented preliminary
measurements the of asymmetry
${\cal A} \equiv {{{\cal B}(b \rightarrow f) -
{\cal B}({\bar b} \rightarrow {\bar f})} \over
{{\cal B}(b \rightarrow f) +
{\cal B}({\bar b} \rightarrow {\bar f})}}$ for five charmless
two-body final states $f$ ($K^- \pi^+$, $K^- \pi^0$, $K^0_s \pi^+$,
$K^- \eta'$, $\omega \pi^+$) \cite{Coan:1999fq}.   Within the Standard Model,
theoretical expectations for these asymmetries range up to $\sim 0.10$
\cite{Ali:1999gb}.  Using the full CLEO~II/II.V data sample, the
statistical precision on ${\cal A}$ is
between $\pm 0.12$ and $\pm 0.25$ for the modes studied.  While these
measurements are not yet a powerful test of the Standard Model, increasing
event samples could render the larger asymmetries measurable within a
few years.

The growing recognition that $B \rightarrow \pi \pi$ will not provide
an easy route to the unitarity triangle parameter $\alpha$ has stimulated
the search for alternatives.  The most promising avenue was suggested 
by Snyder and Quinn \cite{Snyder:1993mx}.  They observe that
a full Dalitz analysis of $B \rightarrow \pi^+ \pi^- \pi^0$ exploits
interference among the different $B \rightarrow \rho \pi$ modes to
remove ambiguities due to unknown phases.  This provides a  
determination of $\alpha$ to within about $6^\circ$ with a sample of
$\sim 1000$ $B \rightarrow \rho \pi$ decays, assuming
the sample to be essentially background-free.

\begin{figure}[tb!]
\begin{center}
\leavevmode
{\epsfxsize=2.6truein \epsfbox{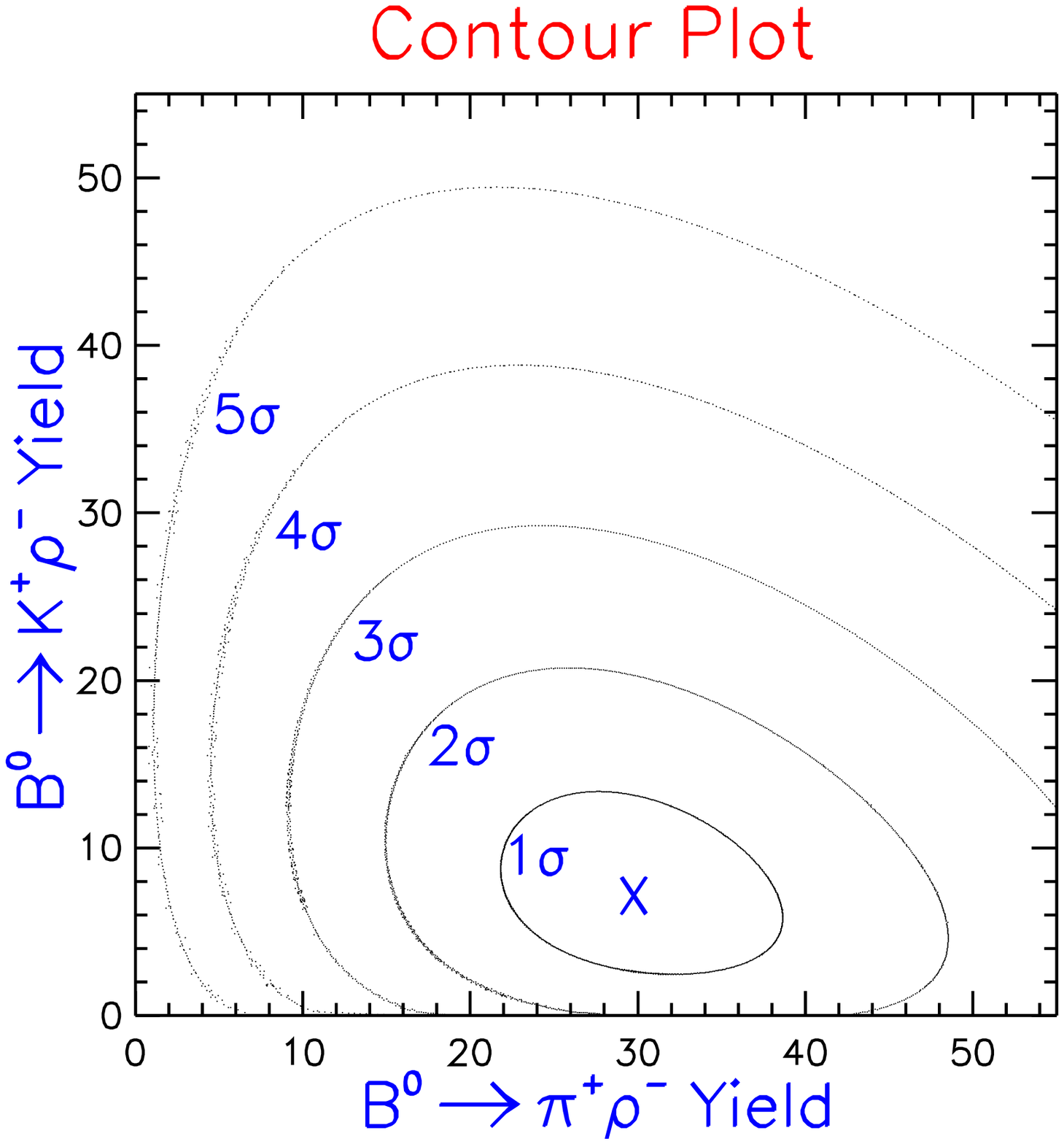}}\hspace*{0.1in}
{\epsfxsize=2.55truein \epsfbox{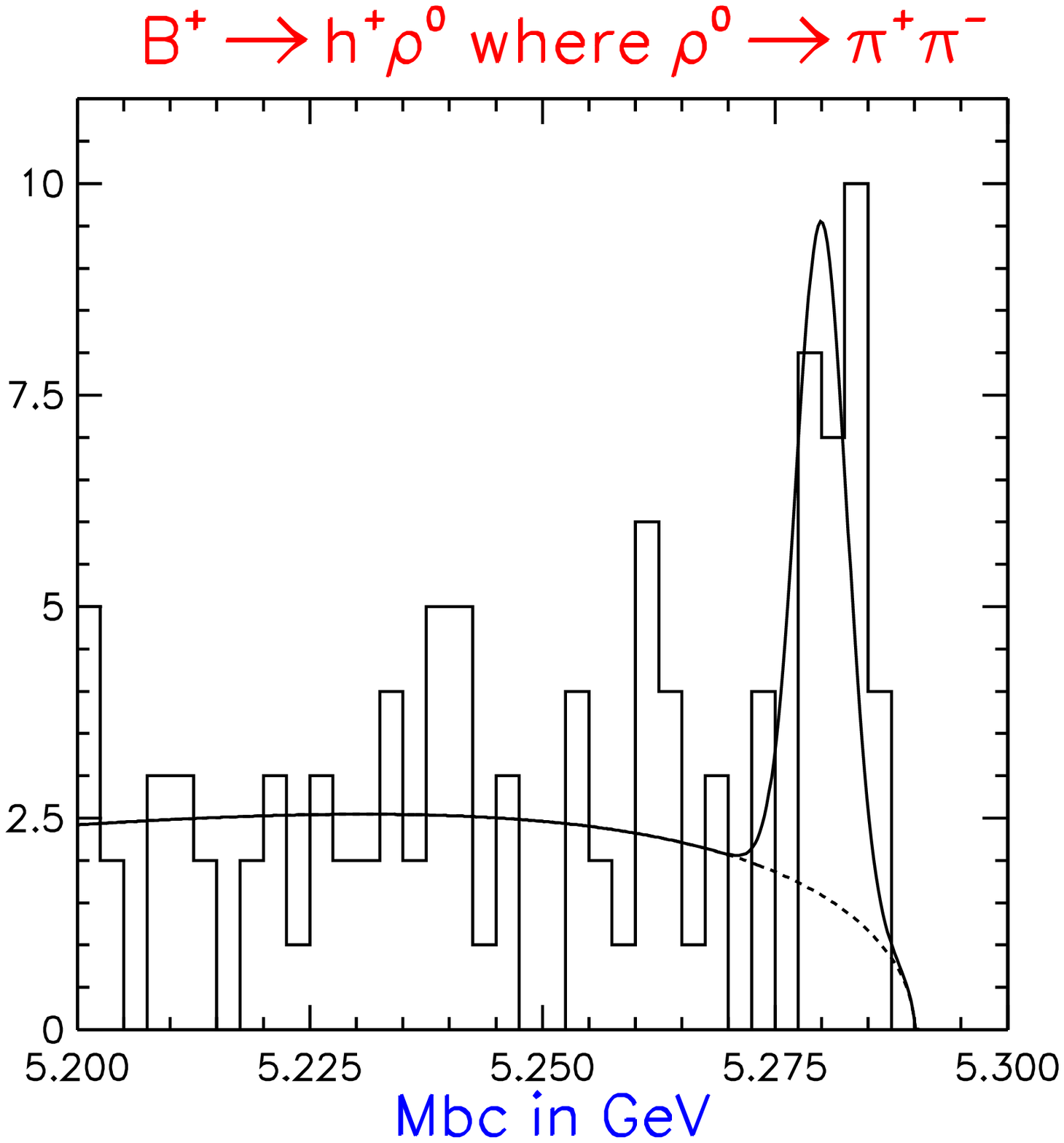}}
\end{center}
\caption{Likelihood contours (left) and projections onto 
beam-constrained mass (right) for $B^+ \rightarrow h^+ \rho^0$
candidates.}
\label{fig:CLEO_rhopi}
\end{figure}

CLEO has presented preliminary results of searches for $B$ decays 
into final states with a $K^*$, $\rho$, $\omega$, or $\phi$ meson and
a second low-mass meson \cite{Gao:1999ik,Bishai:1999x}.  The results for 
all modes investigated are summarized in Table~\ref{tab:CLEO_PV_results}.
\begin{table}[htbp!]
\caption{Results of searches for decays 
$B \rightarrow PV$.  Reconstruction efficiencies ($\epsilon$)
and total detection efficiencies including secondary branching fractions
($\epsilon {\cal B}_s$) are shown, as are the statistical significance,
branching fractions (${\cal B}$) and/or upper limits. }
\begin{center}
\begin{tabular}{lrrrccc}
Final state & Yield(events) & $\epsilon$(\%) & $\epsilon\calB_s$(\%) &
Signif. & \calB($10^{-6})$ & 90\% UL($10^{-6}$) \\ \hline
\omegapi      & $28.5^{+8.2}_{-7.3}$ & 29 & 26 & 6.2 &
     $11.3^{+3.3}_{-2.9} \pm 1.5$ & 17   \\
\omegapiz     & $1.5^{+3.5}_{-1.5}$  & 22 & 19 & 0.6 &
     $0.8^{+1.9}_{-0.8} \pm 0.5$ & 5.8  \\
\omegak       & $7.9^{+6.0}_{-4.7}$  & 29 & 26 & 2.1 &
     $3.2^{+2.4}_{-1.9} \pm 0.8$ & 8.0  \\
\omegakz      & $7.0^{+3.8}_{-2.9}$  & 24 &  7.4 & 3.9 &
     $10.0^{+5.4}_{-4.2} \pm 1.5$ & 21   \\
\omegah       & $35.6^{+8.9}_{-8.0}$ & 29 & 26 & 7.3 &
     $14.3^{+3.6}_{-3.2} \pm 2.1$ &  21  \\
\omegarhop    & $10.8^{+6.6}_{-5.3}$ & 7.1 &  6.3 & 2.8 &
     $18^{+11}_{-9} \pm 6$ & 47   \\
\omegarhoz    & $3.7^{+6.0}_{-3.7}$  & 18 & 16 & 0.9 &
     $0.0^{+5.7 \; +2.9}_{-0.0 \; -0.0}$ & 11  \\
\omegakstp   & $1.0^{+3.6}_{-1.0}$  & 6.8 &  2.0 & 0.3 &
     $5^{+19}_{-5} \pm 6$ & 52   \\
\omegakstz   & $7.0^{+5.2}_{-3.9}$  & 14 & 8.3 & 2.3 &
     $9.1^{+6.7}_{-5.1} \pm 1.9$ & 19   \\
 & & & & & & \\
\rhozpi       & $26.1^{+9.1}_{-8.0}$ & 30 & 30   &  5.2 &
     $15^{+5}_{-5} \pm 4$ &   \\
\rhompi       & $28.5^{+8.9}_{-7.9}$ & 12 & 12   &  5.6 &
     $35^{+11}_{-10} \pm 5$ &   \\
\rhozpiz      & $3.4^{+5.2}_{-3.4}$  & 34 & 34 &     &
                            & 5.1  \\
\rhozk        & $14.8^{+8.8}_{-7.7}$ & 28 & 28   &      &
                          & 22 \\
\rhomk        & $8.3^{+6.3}_{-5.0}$  & 11 & 11   &      &
                          & 25 \\
\rhozkz       & $8.2^{+4.9}_{-3.9}$  &    & 10   &  2.7 &
                          & 27 \\
 & & & & & & \\
\kstzpi       & $12.3^{+5.7}_{-4.7}$ &    & 18   &      &
                          & 27 \\
\kstzpiz      & $0.1^{+2.8}_{-0.1}$  & 37 & 24   &     &
                                  & 4.2 \\
\kstzkkp      & $0.0^{+2.1}_{-0.0}$ &    &  18  &      &
                                  &  12  \\
 & & & & & & \\
\kstppikz     & $10.8^{+4.3}_{-3.5}$ &    &  7   &  5.2 &
     $23^{+9}_{-7}\pm{3}$ &   \\
\kstppikp     & $5.7^{+4.3}_{-3.2}$ &    & 4.1   &  2.5 &
     $20^{+15 \; +3}_{-11 \; -4}$ &   \\
~~\kstppi       &                     &    &       &  5.9 &
     $22^{+8 \; +4}_{-6 \; -5}$ &   \\
 & & & & & & \\
\kstpkkz      & $0.0^{+0.9}_{-0.0}$ &    &  7   &  0.0 &
                                 &  8  \\
\kstpkkp      & $0.0^{+1.3}_{-0.0}$ &    &  4.1   &  0.0 &
                                  &  17  \\
~~\kstpk        &                  &    &      &  0.0 &
                                  &  6  \\
 & & & & & & \\
\phipi        &                      & 54 &  27  &  0.0 &
                                 &  4.0 \\
\phipiz       &                      & 35 &  17  &  0.0 &
                                 &  5.4 \\
\phik         & $2.4^{+3.0}_{-1.9}$  & 53 & 26   &  1.3 &
     $1.6^{+1.9}_{-1.2} \pm 0.2$ &  5.9 \\
\phikz        & $4.3^{+3.2}_{-2.3}$  & 41 & 7.0  &  2.6 &
     $10.7^{+7.8}_{-5.7} \pm 1.1$ & 28  \\
\end{tabular}
\end{center}
\label{tab:CLEO_PV_results}
\end{table}
Fig.~\ref{fig:CLEO_rhopi} shows the likelihood contours and
beam-constrained mass distributions for 
$B^+ \rightarrow \rho^0 h^+$ candidates.
there is clear evidence of a signal, which translates into a
branching fraction measurement 
${\cal B}(B^+ \rightarrow \rho^0 \pi^+)=
(1.5 \pm 0.5 \pm 0.4) \times 10^{-5}$.
There is no significant signal for $B^+ \rightarrow \rho^0 K^+$,
with  a 90\% confidence-level upper limit of
${\cal B}(B^+ \rightarrow \rho^0 K^+)<
2.2 \times 10^{-5}$.  The situation is similar for
$B^0 \rightarrow \rho^\pm h^\mp$, for which the significant 
$B^0 \rightarrow K^{*+} \pi^-$ background demands a cut on the helicity
angle.  Again there is a measurement for the $\rho \pi$ mode
(${\cal B}(B^0 \rightarrow \rho^\pm \pi^\mp)=
(3.5^{+1.1}_{-1.0}\pm 0.5) \times 10^{-5}$), and only an upper limit for 
$\rho K$ (${\cal B}(B^0 \rightarrow  \rho^\pm K^\mp) < 2.5 \times 10^{-5}$
at 90\% confidence level). 

These measurements allow us to assess the feasibility of
measuring $\alpha$ with $B \rightarrow \rho \pi$.  More than 
100~fb$^{-1}$ will be needed to obtain the specified 1000
events.  This sample will require several years of an
asymmetric $B$ factory to accumulate, and the need to reduce and
understand backgrounds will be a major challenge.  

\begin{figure}[tb!]
\begin{center}
\leavevmode
{\epsfxsize=2.8truein \epsfbox{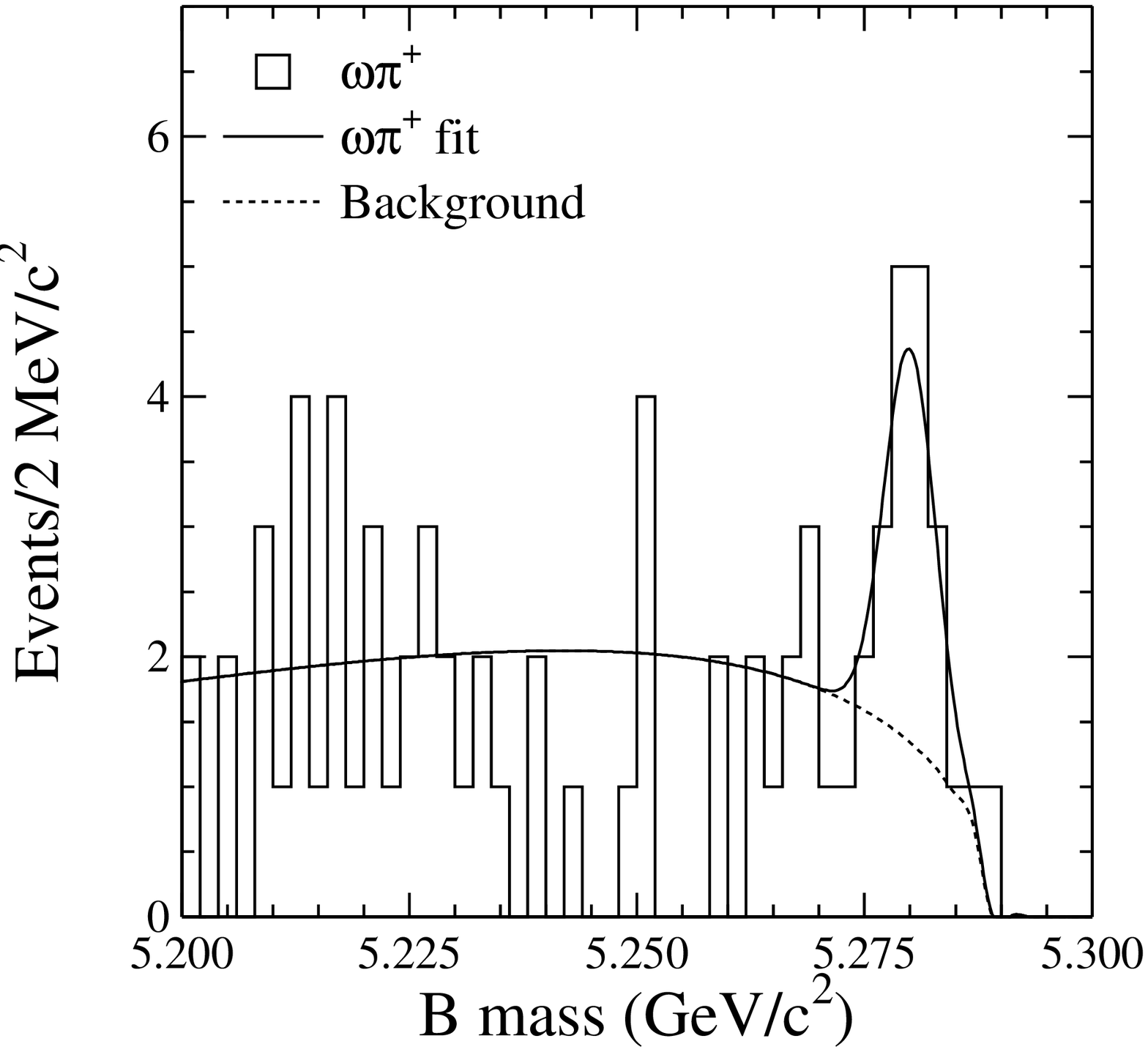}}\hspace*{0.01in}
{\epsfxsize=2.8truein \epsfbox{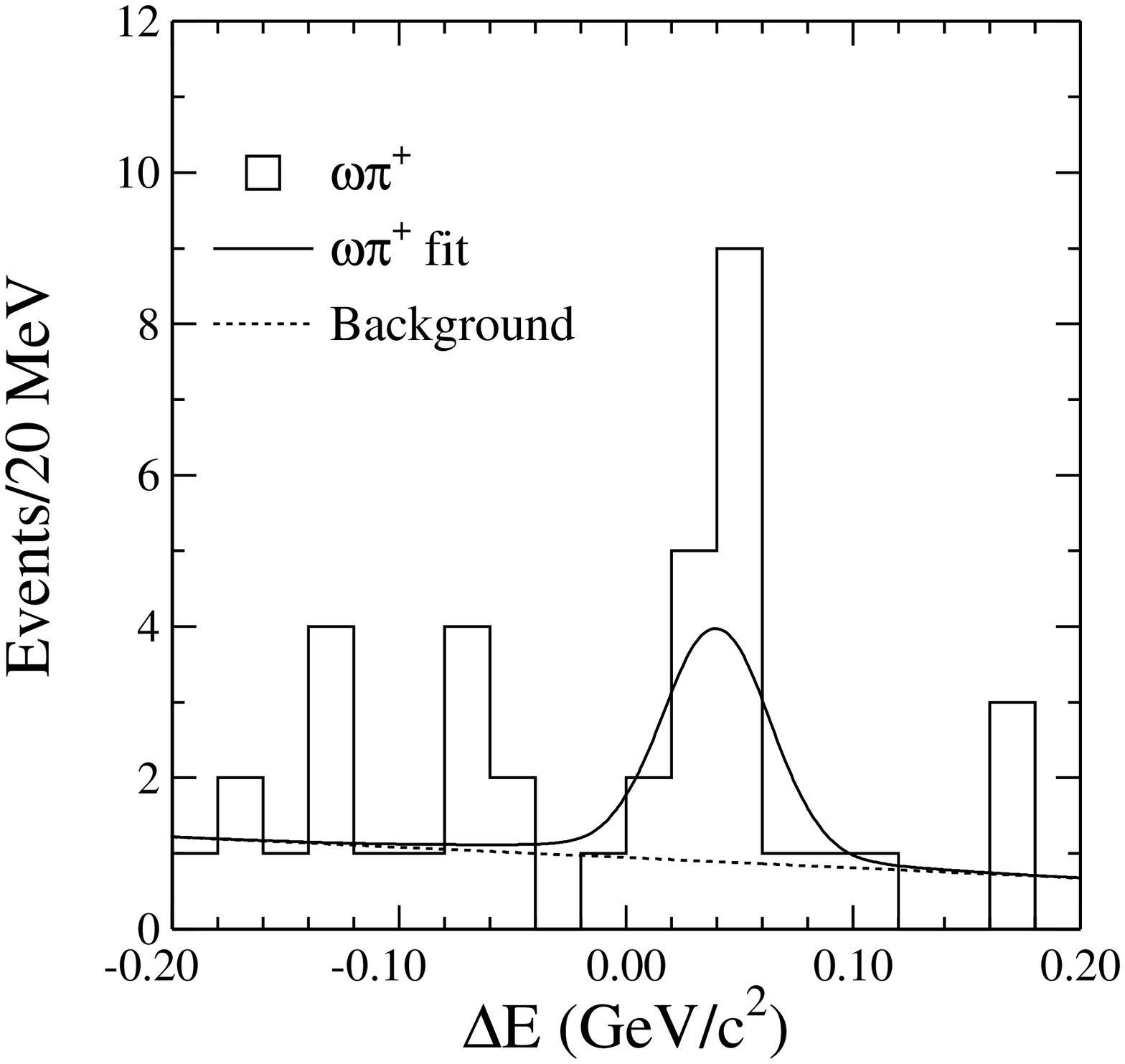}}
\end{center}
\caption{Projections onto beam-constrained mass and
$\Delta E$ for $B \rightarrow \omega \pi^+$.}
\label{fig:CLEO_omegapi}
\end{figure}

Among the other measurements reported in Ref.~\cite{Bishai:1999x}
is the intriguing observation of the decay $B^+ \rightarrow \omega \pi^+$.  
Fig.~\ref{fig:CLEO_omegapi} 
shows the
projected distributions of beam-constrained mass and $\Delta E$.
The signal is solid, and the measured branching fraction 
${\cal B}(B^+ \rightarrow \omega \pi^+)=(11.3^{+3.3}_{-2.9} \pm 1.5) \times 10^{-6}$ 
agrees well with CLEO's measurement for 
$B \rightarrow \rho^0 \pi^+$, as expected from isospin.
While the branching fraction for $B^+ \rightarrow \omega \pi^+$ is consistent
with CLEO's published upper limit on this 
mode \cite{Bergfeld:1998ik}, the new upper limit on $B \rightarrow \omega K^+$
conflicts with the previously reported observation \cite{Bergfeld:1998ik}.  
There is no obvious explanation for this change other than a fluctuation
in the previous search.  The new measurement is an improvement over the
first in several ways.  The data sample has almost tripled and 
analysis-procedure improvements have increased the 
reconstruction efficiencies by between 10\% and 20\%.  

\begin{figure}[tb!]
\begin{center}
\epsfig{file=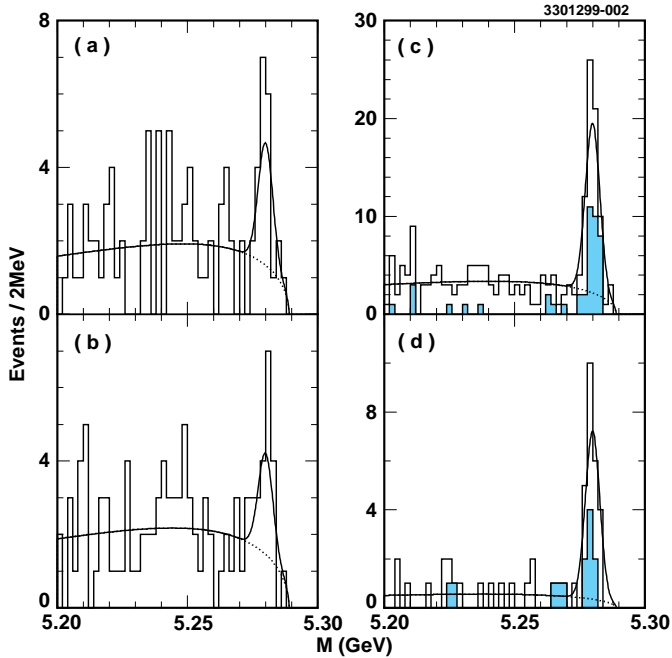,width=3.5in}
\caption{Projections onto beam-constrained mass for (a) 
$B \rightarrow \eta K^{*+}$, (b) $B \rightarrow \eta K^{*0}$,
(c) $B \rightarrow \eta' K^+$, and (d) $B \rightarrow \eta' K^0$.
The shaded histograms in (c) and (d) correspond to  
$\eta' \rightarrow \eta \pi \pi$, $\eta \rightarrow \gamma \gamma$, 
while the unshaded histograms are 
$\eta' \rightarrow \rho \gamma$.  The solid (dashed) lines show the
projections of the full fit (background only).}
\label{fig:CLEO_eta_1}
\end{center}
\end{figure}

A ``poster child'' for the challenge of interpreting charmless hadronic
$B$ decays is the decay $B \rightarrow \eta' K$.  In 1998
CLEO reported an unexpectedly large branching fraction for this 
mode \cite{Behrens:1998dn}, stimulating considerable theoretical
interest.  An updated search for two-body $B$ decays to $\eta$ and
$\eta'$ has now been reported \cite{Richichi:1999kj}, and
the mystery has not gone away.  The distributions of beam-constrained 
mass for $B \rightarrow \eta K^{*}$ and 
$B \rightarrow \eta' K$ are shown in Fig.~\ref{fig:CLEO_eta_1}.  The 
branching fractions for the modes with clear signals are 
${\cal B}(B^+ \rightarrow \eta' K^+ = (6.5^{+1.5}_{-1.4} \pm 0.9)
\times 10^{-5}$ and
${\cal B}(B^0 \rightarrow \eta' K^0 = (4.7^{+2.7}_{-1.4} \pm 0.9)
\times 10^{-5}$.  These exceed all theoretical 
predictions \cite{chau,du,kps}.  No statistically significant
signal is observed among the 17 other modes involving $\eta$
and $\eta'$.  A few of these limits are impinging on the
predicted Standard Model range.

\begin{figure}[tb!]
\begin{center}
\leavevmode
{\epsfxsize=2.95truein \epsfbox{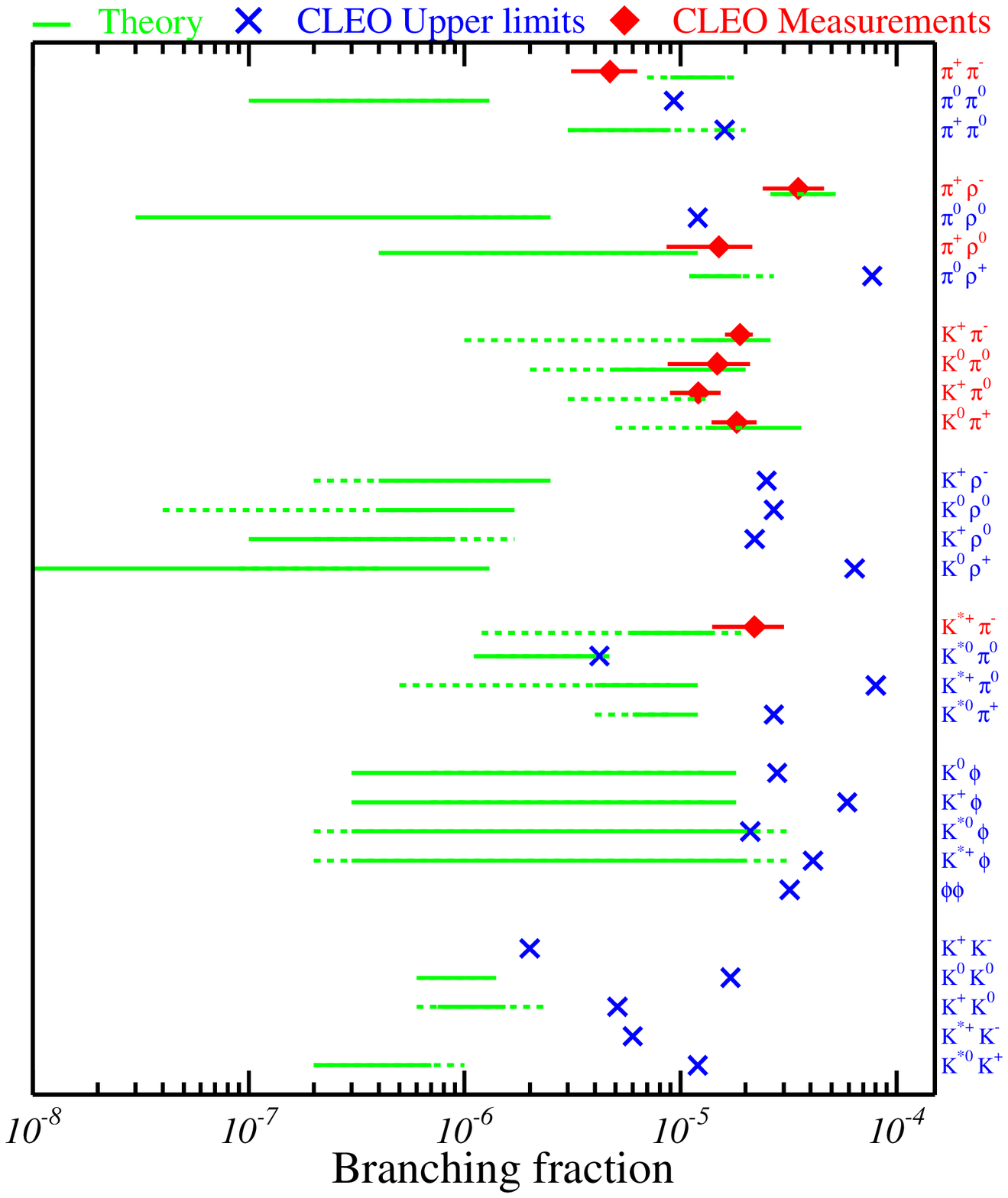}}\hspace*{0.01in}
{\epsfxsize=2.95truein \epsfbox{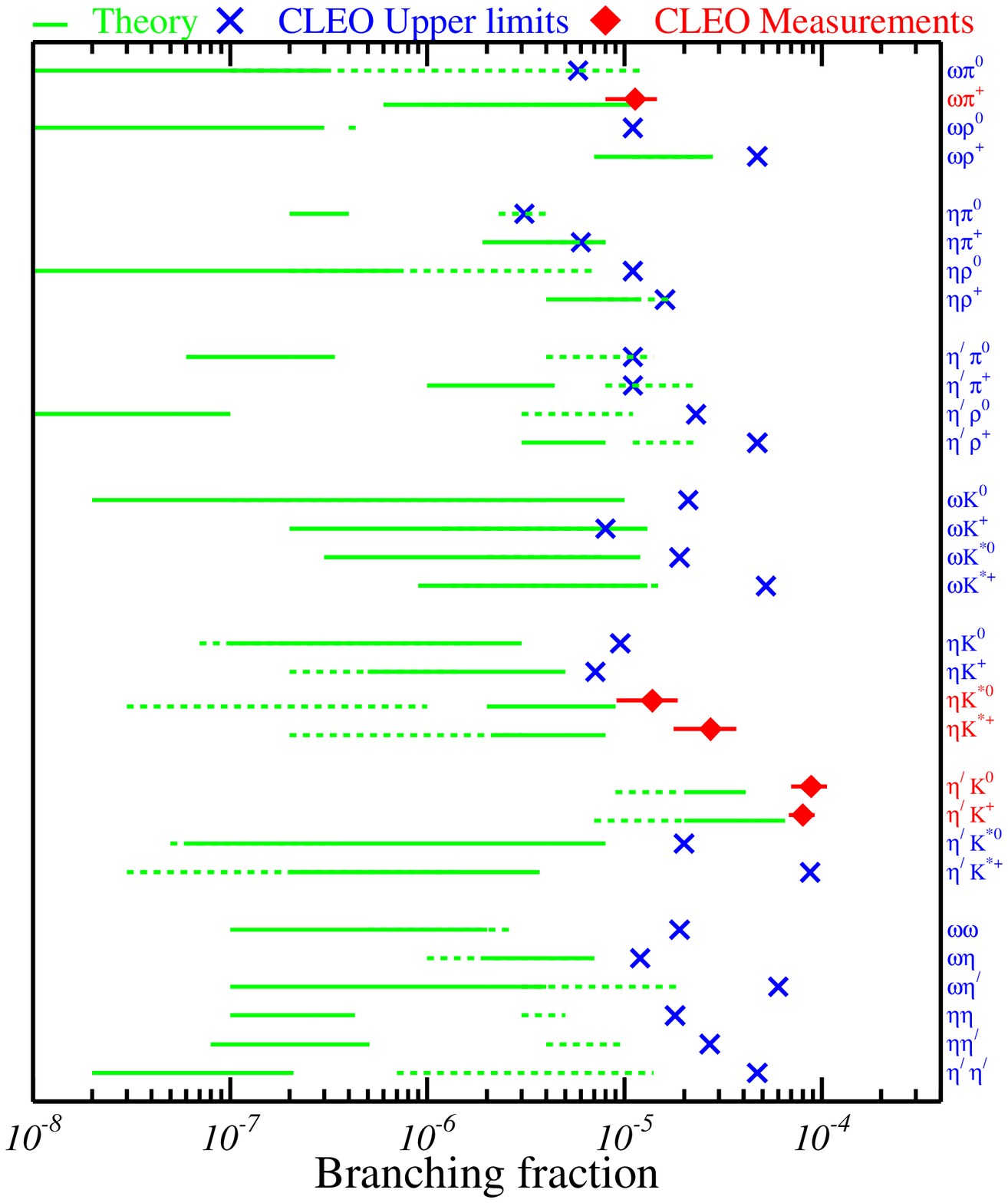}}
\end{center}
\caption{Summary graphs of CLEO rare-$B$ measurements (left) for 
$B \rightarrow K \pi$ and (right) for other rare modes.  The points 
with errors represent measured branching fractions for modes with
definite observations, while the crosses are 90\% confidence-level
upper limits.  The lines give the range of theoretical predictions.}
\label{fig:CLEO_sum_rare}
\end{figure}

Fig.~\ref{fig:CLEO_sum_rare} shows summary graphs for all CLEO-measured
rare two-body $B$-decay processes.  Comparisons with theoretical predictions
are included.  Perhaps the most
impressive feature of the work done is the breadth of the set of modes
that have been measured.  This prepares us for global analyses of rare 
charmless hadronic decays in which multiple  measurements of related 
modes are used to extract detailed information about 
the CKM matrix.  I return to this question in Sec.~\ref{sec:CKM}.

\subsection{$b \rightarrow s \gamma$ and $b \rightarrow s \ell^+ \ell^-$}

\begin{figure}[tb!]
\begin{center}
\leavevmode
{\epsfxsize=2.5truein \epsfysize=2.5truein \epsfbox{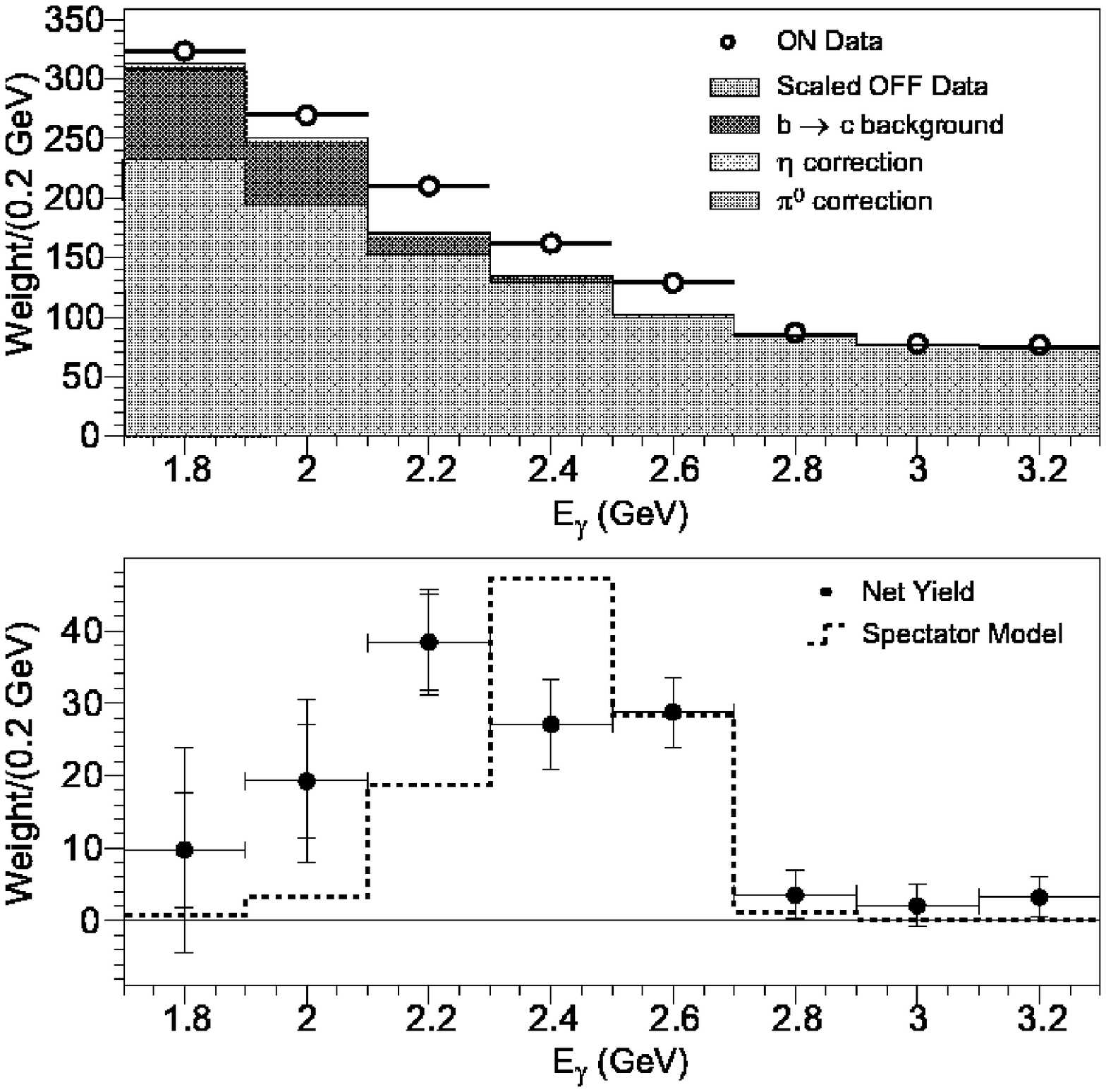}}\hspace*{0.15in}
{\epsfxsize=2.5truein \epsfysize=2.5truein \epsfbox{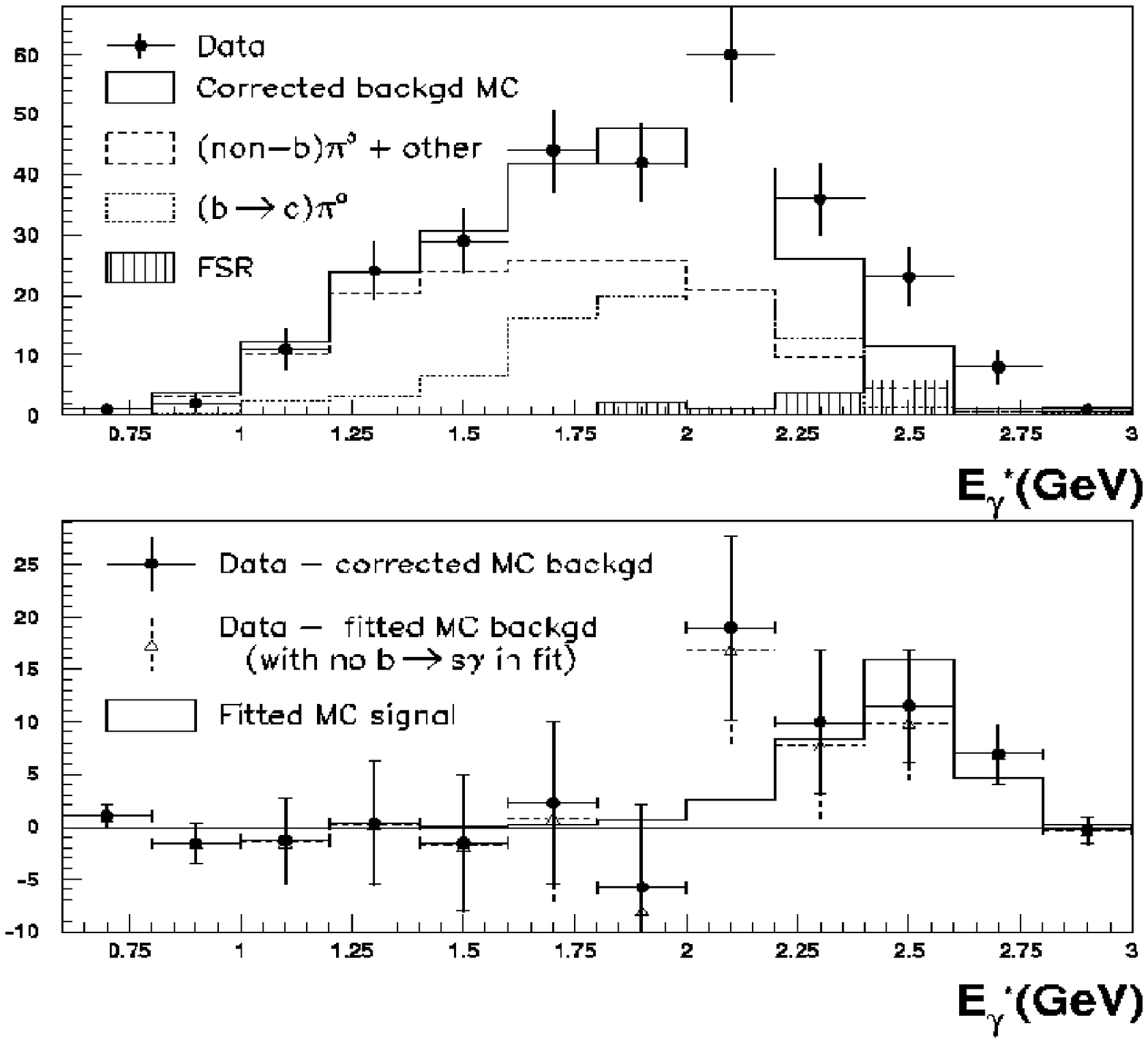}}
\end{center}
\caption{Left: CLEO inclusive photon spectrum (top) for On- and 
Below-$\Upsilon(4S)$ data, and (bottom) 
fully background-subtracted.  Right: ALEPH photon spectrum, 
measured in the rest frame of the reconstructed jet, both raw and 
background-subtracted.}
\label{fig:bsgamma}
\end{figure}

Inclusive measurements of $b \rightarrow s \gamma$
provide powerful constraints on physics beyond the Standard Model.
CLEO has recently presented an updated analysis of 3.3 million
$B {\bar B}$ events \cite{Ahmed:1999fh}.  The technique is an amalgam 
of continuum suppression through shape variables with a neural net and 
pseudo-reconstruction of $B \rightarrow X_s \gamma$.  For the latter,
the $X_s$ consists of a charged or neutral kaon
and up to four pions, one of which can be a $\pi^0$.  The photon spectrum
is shown in Fig.~\ref{fig:bsgamma}, and the branching fraction measurement 
is ${\cal B} = (3.15 \pm 0.35 \pm 0.32 \pm 0.26) \times 10^{-4}$,
where the errors are statistical, systematic and model-dependent,
respectively.  ALEPH has also presented an inclusive measurement of 
$b \rightarrow s \gamma$ \cite{Barate:1998vz}.  In their analysis
non-$B$ backgrounds are suppressed with an opposite-hemisphere
lifetime tag.  As for CLEO a pseudo-reconstruction approach is
employed, in which $B \rightarrow X_s \gamma$ is assembled 
from between one and eight tracks, $K^0_s$'s and $\pi^0$'s.
The photon spectrum is shown in Fig.~\ref{fig:bsgamma}.  ALEPH's
result is ${\cal B} = (3.11 \pm 0.80 \pm 0.72) \times 10^{-4}$,
very consistent with CLEO's.

It has recently been recognized that additional sensitivity to new
physics is provided by the rate asymmetry
${\cal A} = {{\Gamma(b \rightarrow s \gamma) -
\Gamma({\bar b} \rightarrow {\bar s} \gamma)} \over
{\Gamma(b \rightarrow s \gamma) +
\Gamma({\bar b} \rightarrow {\bar s} \gamma)}}$.
Some non-Standard Model predictions give asymmetries as large as
~40\% \cite{Kagan:1998bh,Aoki:1999kf}.
CLEO's updated study of inclusive 
$b \rightarrow s \gamma$ \cite{Ahmed:1999fh} includes an extension
of the pseudo-reconstruction analysis to measure this asymmetry.  The
strangeness content of the $X_s$ system can be used to tag the flavor 
of the decaying $B$, but mistags and untaggable states must be 
carefully accounted for.  CLEO finds
${\cal A} = (0.16 \pm 0.14 \pm 0.05) \times (1.0 \pm 0.14)$,
with both additive and multiplicative (mistagging rate) systematic 
uncertainties.  The 90\% confidence-level range on ${\cal A}$
is $-0.09 < {\cal A} < 0.42$.

\begin{figure}[tb!]
\begin{center}
\leavevmode
{\epsfxsize=2.95truein \epsfbox{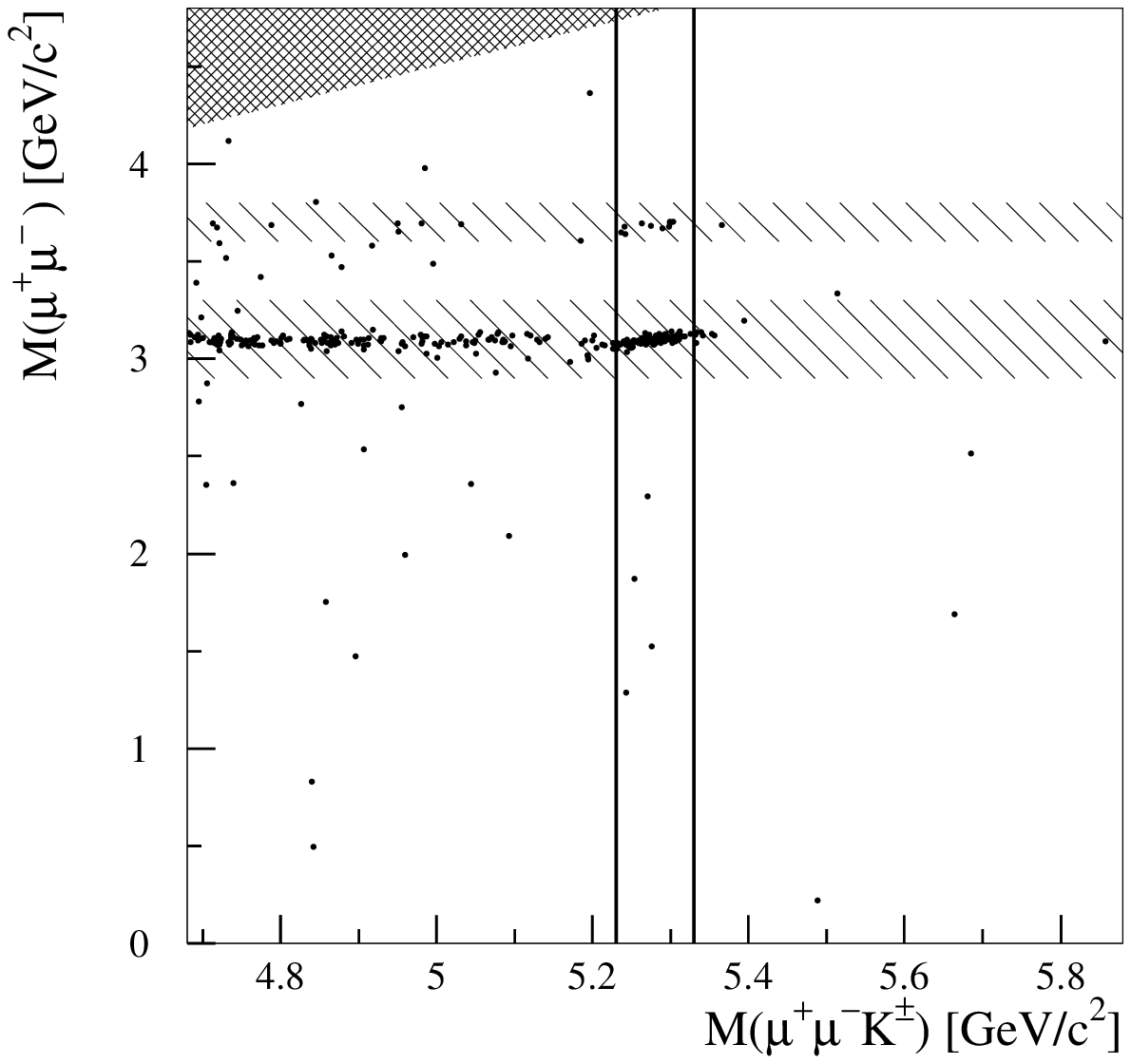}}\hspace*{0.1in}
{\epsfxsize=2.95truein \epsfbox{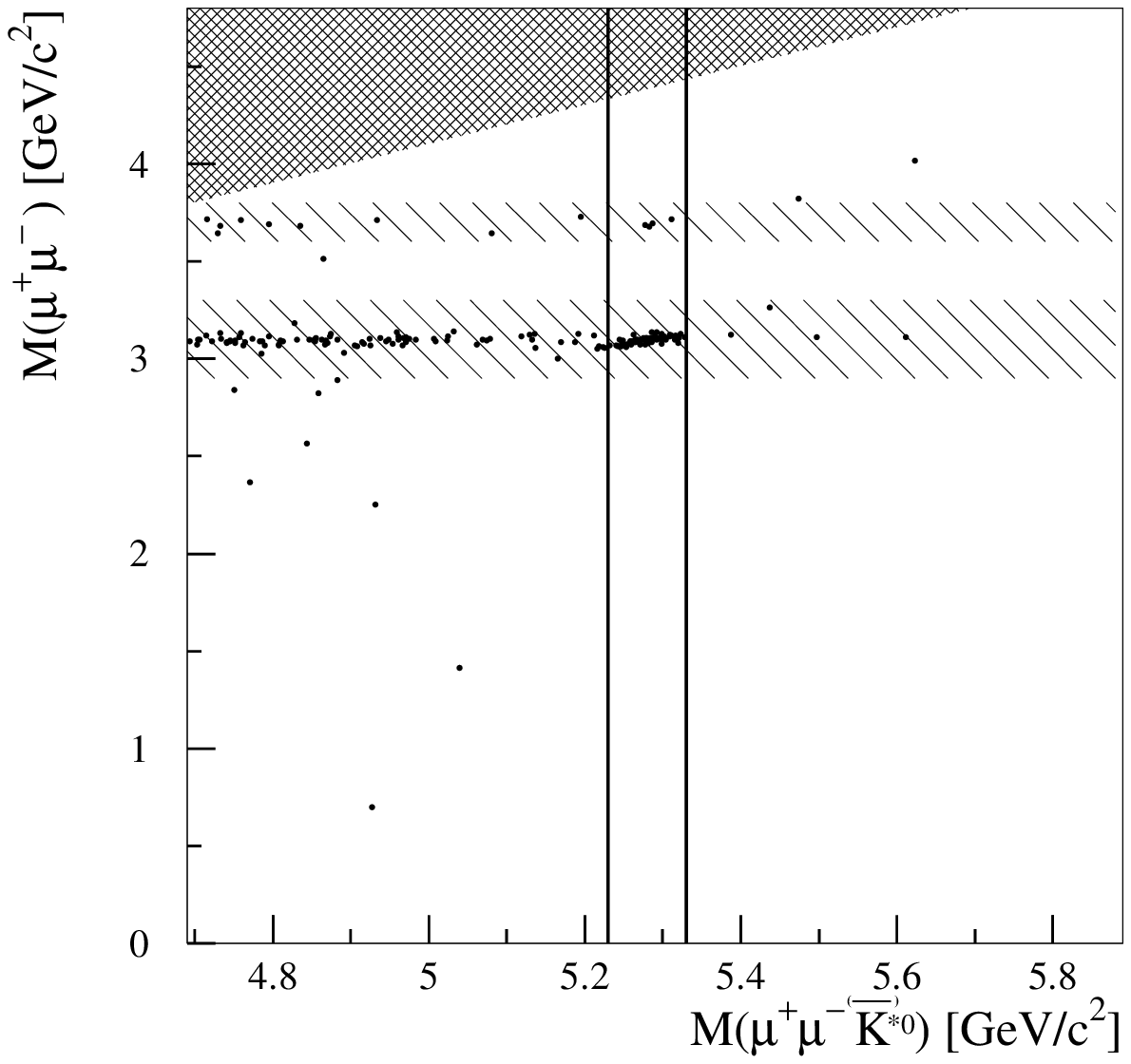}}
\end{center}
\caption{Plots of $\mu^+ \mu^-$ mass vs. $B$-candidate
mass for (left) $B \rightarrow \mu^+ \mu^- K^\pm$ and (right)
$B^0 \rightarrow \mu^+ \mu^- K^{*0}$.  Cross-hatched bands are
excluded charmonium regions.}
\label{fig:CDF_ll}
\end{figure}

A probe of non-Standard Model physics similar to $b \rightarrow s \gamma$
is provided by $b \rightarrow s \ell^+ \ell^-$.  CDF dominates the
search for the exclusive decays 
$B \rightarrow K/K^* \mu^+ \mu^-$ \cite{Affolder:1999eb}, 
with a very large sample
of hadronically produced $B$'s and the capability to tag $B$ production
by displaced vertices.
Fig.~\ref{fig:CDF_ll} shows the distributions of $M(\mu^+ \mu^-)$ vs.
$M(K/K^* \mu^+ \mu^-$) for the CDF data.  The background is largely
confined to the easily excluded $J/\psi$ and $\psi'$ dilepton mass
bands, leaving a very clean measurement.  CDF obtains the 90\%
confidence limits ${\cal B}(B^+ \rightarrow K^+ \mu^+ \mu^-)~<~
5.2 \times 10^{-6}$ (Standard Model prediction: $0.3-0.7 \times 10^{-6}$),
and ${\cal B}(B^0 \rightarrow K^{*0} \mu^+ \mu^-)~<~4.0 \times 10^{-6}$
(Standard Model: $1-4 \times 10^{-6}$).  With a data sample
of 2~$fb^{-1}$ expected for Run~II, the observation of this mode should
not be far off.

Since interpretation of the exclusive decays is somewhat problematic
(as for $b \rightarrow s \gamma$), inclusive measurements
would have some advantage.  Both D0 \cite{Abbott:1998hc}
and CLEO \cite{Glenn:1998gh} have reported inclusive analyses.  For
D0 this is a search for lepton pairs with masses between the charmonium
region and the $B$ meson, a window that includes only a portion
of $b \rightarrow s \ell^+ \ell^-$, but which is very clean.  CLEO
employs a pseudo-reconstruction technique similar to the 
$b \rightarrow s \gamma$ procedure.  All limits obtained are an
order of magnitude or more above Standard Model expectation.

\section{Interpretation -- CKM}
\label{sec:CKM}

Information relevant to the determination of the CKM parameters is
being accumulated at an accelerating rate.  While principal 
responsibility for its interpretation in these proceedings falls to 
Adam Falk \cite{Falk:1999sp}, I will briefly comment on the conventional 
view and then highlight a speculative
interpretation of CLEO's rare-$B$-decay data.

\begin{figure}[tb!]
\begin{center}
\leavevmode
{\epsfxsize=2.7truein \epsfysize=1.95truein \epsfbox{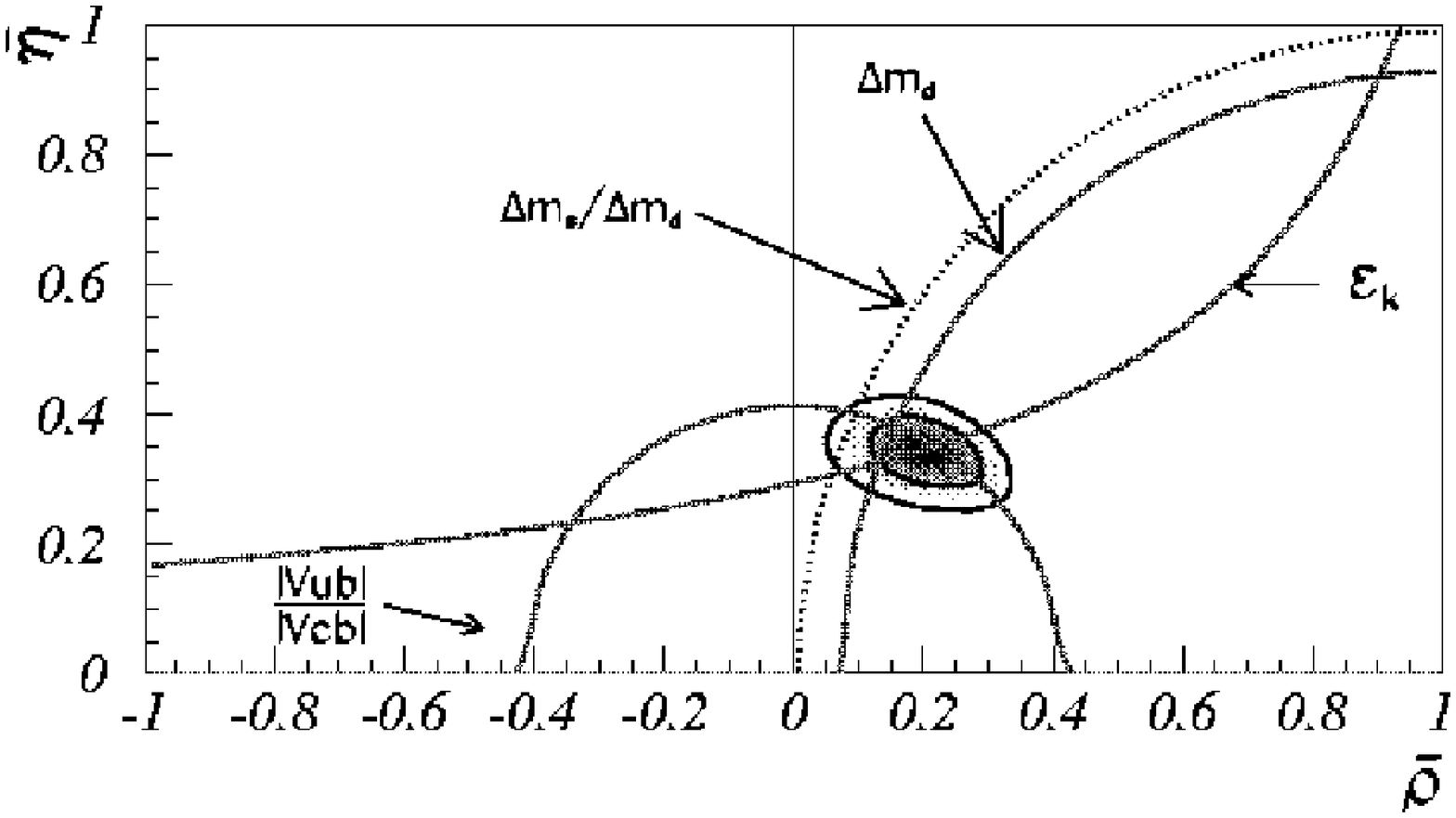}}\hspace*{0.05in}
{\epsfxsize=2.9truein \epsfysize=2.05truein \epsfbox{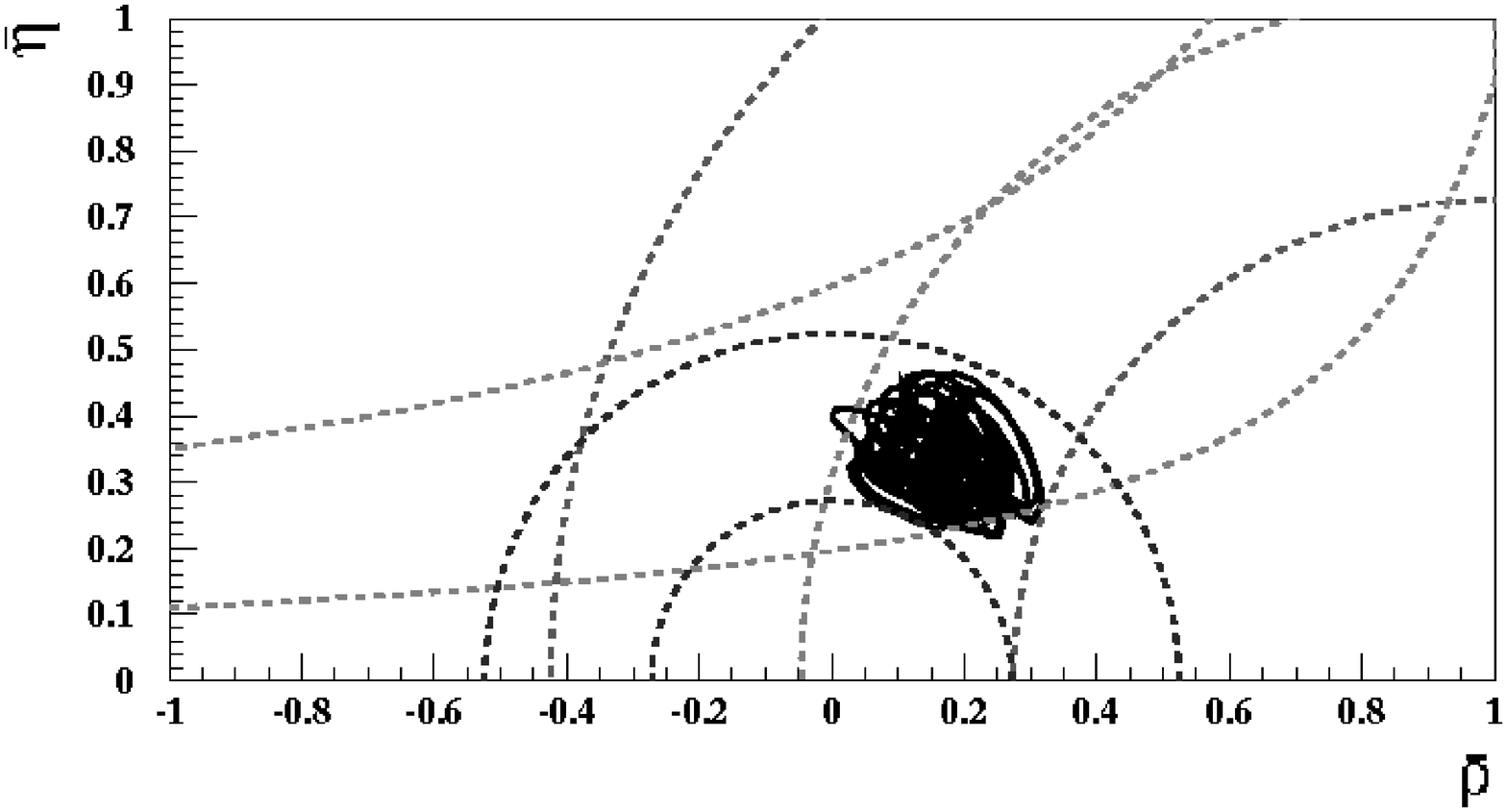}}
\end{center}
\caption{Left: Allowed region for ${\bar \rho}$ and ${\bar \eta}$
from Parodi {\it et al.} \cite{Parodi:1999nr}, with
68\% and 95\% confidence-level contours.
Right: Allowed region according
to Plaszczynski \cite{Plaszczynski:1999}.  Small
contours represents different theoretical models, with the envelope giving
the overall 95\% confidence-level range.  The curves are the
usual experimental constraints.}
\label{fig:CKM_fits}
\end{figure}

A number of authors have incorporated the principal constraints
from $B$ decay ($|V_{ub}/V_{cb}|$, $\Delta m_d$ and the limit on
$\Delta m_s$) with input from $K^0_L$ CP violation ($|\epsilon_K|$)
in global fits to obtain the Wolfenstein parameters ${\bar \rho}$
and ${\bar \eta}$, and the angles $\alpha$,
$\beta$ and $\gamma$ of the unitarity triangle.
Parodi {\it et al.} \cite{Parodi:1999nr} and Mele \cite{Mele:1999bf}
have performed maximum likelihood fits that assign Gaussian errors
to several theoretical inputs.  The fits of Parodi {\it et al.}
give the solution shown on the left-hand side of Fig.~\ref{fig:CKM_fits},
leading to ${\bar \rho} = 0.202^{+0.053}_{-0.059}$ and
${\bar \eta} = 0.340 \pm 0.035$, which in turn give
${\rm sin} 2\alpha = -0.26^{+0.29}_{-0.28}$,
${\rm sin} 2\beta = 0.725^{+0.050}_{-0.060}$ and
$\gamma = (59.5^{+8.5}_{-7.5})$ degrees.

Stone has pointed out the danger of 
underestimating the overall uncertainty when 
assuming Gaussian errors for theoretical inputs \cite{Stone:1999td}.
Plaszczynski \cite{Plaszczynski:1999} has taken a much more cautious
approach, considering all theoretical models on an equal basis
and presenting the full spread in the resulting parameter values
as shown in Fig.~\ref{fig:CKM_fits}.  He
obtains the larger ranges
$0 \le {\bar \rho} \le 0.3$ and $0.2 \le {\bar \eta} \le 0.45$,
leading to $0.50 \le {\rm sin} 2\beta \le 0.85$ and
$-0.95 \le {\rm sin} 2\alpha \le 0.50$.  Of course everything is 
consistent with CDF's first direct measurement, 
${\rm sin} 2\beta = 0.79^{+0.41}_{-0.44}$ \cite{Bortoletto:1999up}.

Because rare hadronic $B$ decays incorporate both penguin and 
$b \rightarrow u$ tree processes, their rates and CP asymmetries carry 
information about weak phases.  In particular, it has been suggested
by several authors that combinations of measured rates can be used
to extract the value of $\gamma$, the phase of $V_{ub}^*$.  The
first suggestions \cite{Fleischer:1998um,Neubert:1998pt,Neubert:1999ua} 
focused on the $B \rightarrow K \pi$ branching fractions, but 
these approaches do not set significant bounds on $\gamma$
with current data.

\begin{figure}[tb!]
\begin{center}
\epsfig{file=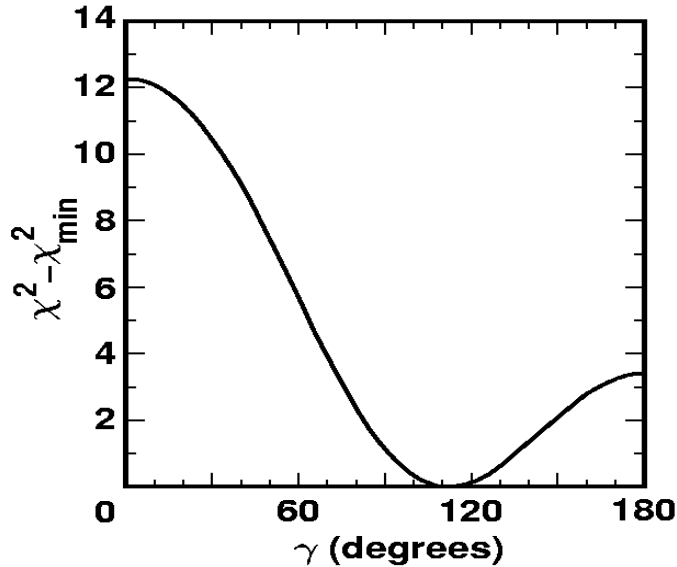,width=3.5in,height=3.in}
\caption{$\chi^2$ vs. $\gamma$ for CLEO's fit to CP-averaged 
charmless $B$ branching fractions.}
\label{fig:CLEO_gamma}
\end{center}
\end{figure}

A much more aggressive procedure to extract maximal information 
from the data has been suggested by Hou, Smith and 
W\"urthwein \cite{Hou:1999qp}.  They assume factorization holds and 
write the $B$-decay amplitudes in terms of five parameters:
$\gamma = Arg(V^*_{ub})$, $|V_{ub}/V_{cb}|$, $R_{su}$ (incorporating
information about quark masses), $F^{B \rightarrow \pi}$ 
($B \rightarrow \pi$ form factor), and $A_0^{B \rightarrow \rho}$
($B \rightarrow \rho$ form factor).
The CP-averaged branching fractions for 14 of CLEO's measured charmless
two-body decays \cite{Kwon:1999hx,Bishai:1999x} were fitted with 
this parameterization.  (Modes with $\eta$ and $\eta'$ were excluded
based on their anomalies.)  The constraint
$|V_{ub}/V_{cb}| = 0.08 \pm 0.02$ was imposed.
The result of the fit is shown in Fig.~\ref{fig:CLEO_gamma}.  The
minimum $\chi^2$ (10.3 for 10 degrees of freedom) occurs for
$\gamma = 113^{+25}_{-23}$ degrees.  The other fit parameters all
give very reasonable values.  This result agrees with earlier
observations that CLEO data favor $\cos \gamma < 0$ 
\cite{He:1999mn,Hou:2000tf}.

This is an intriguing result.  The reasonable values of the
fit parameters other than $\gamma$ suggest that there may be some
validity, in spite of the very model-dependent
assumptions.  On the other hand, this may be nothing more than
a misleading coincidence.  Skepticism is appropriate.

\section{Other Topics}
\label{sec:Other}

It is impossible to report exhaustively on all of the activity
in heavy-quark decays within a single review.  While my focus has been
on CKM tests and measurements relevant to CP violation, there is 
other work that is also having impact.  Even in the $B$ sector 
I have had to ignore some work, including CLEO and LEP studies of 
hadronic decays and exclusive semileptonic decays to charm, that are 
important elements of a comprehensive understanding of $B$ decay.

Charm physics remains an extremely valuable complement to $b$ physics 
in our program of Standard Model testing.  Because the expected rates
for rare FCNC processes are extremely small, the potential to see 
new physics in $D {\bar D}$ mixing, in rare $D$-meson decays or in 
CP-violating processes is great.  Additionally, studies of semileptonic 
and leptonic charm decays are an important adjunct to the CKM measurements, 
with potential to reduce model uncertainties in the extraction of 
$V_{ub}$ and other parameters.
A number of experiments have presented
new results on charm decays, with much more on the way.
This is a very broad program, the components of which
have been the subject of several excellent recent reviews, including
that of lifetimes and mixing elsewhere in these 
proceedings \cite{Blaylock:1999pz}.  

As in $B$ physics, studies of hadronic charm decays 
\cite{Meadows:HF8} are important for
developing a comprehensive understanding of heavy flavors, probing
questions of final-state interactions and interference
effects.  Both meson and baryon decays are useful in this effort,
and previously reported results from E791, CLEO and other experiments
will be greatly enhanced by FOCUS and SELEX. 

A number of new form-factor measurements for semileptonic $D$ and
$D_s$ decays have been presented in the past year by E687 \cite{Stanton:HF8}.
FOCUS will soon have results from larger samples, and CLEO will also
extend previous studies to their full data set.  In tandem with HQET 
these measurements will significantly reduce
model uncertainties in the extraction of $|V_{ub}|$
from data on semileptonic $B$ decays.  Measurements of heavy-meson
decay constants in leptonic decays of charmed mesons provide input
to $B$-physics analyses and tests of lattice calculations.

New limits on rare or forbidden charm decays \cite{PSheldon:HF8} have
been presented by E791.  A blind search for 24 modes was performed,
with no signals observed in any and 90\% confidence-level upper limits
that range from $\sim 10^{-3}$ or $10^{-4}$ for $K/\pi \ell^+ \ell^-$
to less than $\sim 10^{-5}$ for $\ell^+ \ell^-$.  Again FOCUS will
benefit from much greater statistics, with improvements in sensitivity
for these modes of an order of magnitude or better.

\section{Summary and Conclusion}
\label{sec:Concl}

The past several years have seen steady progress on a broad program
of Standard Model tests in $B$ decays, but there remains much to be done.  

The embarrassment of the $Z^0/\Upsilon(4S)$ disagreement on the
$B$ semileptonic branching fraction has eased.  The basic
experimental observation that there are too few semileptonic decays 
for the observed multiplicity of charm quarks is still with us, but
it is not of crisis proportions.  Theoretical tools for describing
semileptonic decays have matured, but underlying assumptions like
quark-hadron duality must be scrutinized.  Hints
of inconsistency between HQET-inspired interpretations of CLEO's
hadronic-mass and lepton-energy moments in semileptonic $B$ decays
are troubling.  A great deal more data and a great deal of work
will be required of to reach final conclusion on the values of 
$V_{ub}$ and $V_{cb}$.  Intensive theory/experiment
collaboration is a big plus.

In rare $B$ decays we have a number of major developments.  The
decay $B \rightarrow \pi \pi$ has been observed, and the
rare hadronic decay picture is filling in with more measurements
and tighter limits.

We stand on the verge of truly powerful tests of the Standard
Model.  First efforts to measure CP asymmetries and CDF's first
measurement of ${\rm sin} 2\beta$ are opening salvoes in the
next phase of the campaign to make redundant measurements of 
the sides and angles of the unitarity triangle.  So far fits to
the usual experimental constraints show the Standard Model to 
be holding up well, but this is only the beginning.

The exciting future of heavy flavor physics is well
documented elsewhere in these proceedings.  The three $e^+e^-$ $B$ 
factories, complemented by the upgraded Tevatron detectors, will 
produce a wealth of new physics.  It is to be hoped that these 
facilities, their successor $e^+e^-$ machines of still higher 
luminosity, and specialized detectors at hadron colliders,
will carry us well beyond the Standard Model.

\def\Discussion{
\setlength{\parskip}{0.3cm}\setlength{\parindent}{0.0cm}
     \bigskip\bigskip      {\Large {\bf Discussion}} \bigskip}
\def\speaker#1{{\bf #1:}\ }




\end{document}